\documentclass[iop,apj,tighten,numberedappendix,twocolappendix]{emulateapj}
\usepackage[breaklinks,colorlinks,urlcolor=blue,citecolor=blue,linkcolor=blue]{hyperref}

\usepackage{amsmath,amssymb}
\usepackage{cleveref}
\usepackage{graphicx}
\usepackage{bm}
\usepackage{xcolor,ulem}
\usepackage[toc,title,page]{appendix}
\usepackage{tocbibind}
\usepackage{color}

\usepackage{wasysym}    
\usepackage{upgreek}    
\usepackage{gensymb}    
\usepackage{orcidlink}  

\newcommand{\orcid}[1]{\href{https://orcid.orffi#1}{\textcolor[HTML]{A6CE39}{\aiOrcid}}}

\begin{document}

\title[Relativistic binary precession]{Relativistic binary precession: impact on eccentric binary accretion and multi-messenger astronomy}

\author{Stanislav DeLaurentiis$^{1,2}$\orcidlink{0000-0002-8922-825X},
Zoltán Haiman$^{2,3}$\orcidlink{0000-0003-3633-5403},
John Ryan Westernacher-Schneider$^{4}$\orcidlink{0000-0002-3047-7200},
Luke Major Krauth$^{2}$\orcidlink{0009-0002-0839-7893},
Jordy Davelaar$^{5,2}$\orcidlink{0000-0002-2685-2434},
Jonathan Zrake$^{6}$\orcidlink{0000-0002-1895-6516},
Andrew MacFadyen$^{7}$\orcidlink{0000-0002-0106-9013}
\\
$^{1}$Department of Applied Mathematics and Theoretical Physics, University of Cambridge, 
Wilberforce Road, Cambridge CB3 0WA, UK\\
$^{2}$Department of Astronomy, Columbia University, 550 W. 120th Street, New York, NY
10027, USA\\
$^{3}$Department of Physics, Columbia University, 550 W. 120th Street, New York, NY
10027, USA\\
$^{4}$Leiden Observatory, Leiden University, P.O. Box 9513, 2300 RA Leiden, The Netherlands\\
$^{5}$Center for Computational Astrophysics, Flatiron Institute, 162 Fifth Avenue, New York, NY 10010, USA\\
$^{6}$Department of Physics and Astronomy, Clemson University, Clemson, SC 29634, USA\\
$^{7}$Center for Cosmology and Particle Physics, Physics Department, New York University, New York, NY 10003, USA}\email{E-mail: sod2112@columbia.edu}

\begin{abstract}
Recent hydrodynamical simulations have shown that circumbinary gas disks drive the orbits of binary black holes to become eccentric, even when general relativistic corrections to the orbit are significant. Here, we study the general relativistic (GR) apsidal precession of eccentric equal-mass binary black holes in circumbinary disks (CBDs) via two-dimensional hydrodynamical simulations. We perform a suite of simulations comparing precessing and non-precessing binaries across a range of eccentricities, semi-major axes, and precession rates. We find that the GR precession of the binary's semi-major axis can introduce a dominant modulation in the binary's accretion rate and the corresponding high-energy electromagnetic light-curves. We discuss the conditions under which this occurs and its detailed characteristics and mechanism. Finally, we discuss the potential to observe these precession signatures in electromagnetic and gravitational wave (GW) observations, as well as the precession signal's unique importance as a potential tool to constrain the mass, eccentricity, and semi-major axis of binary merger events.
\end{abstract}




\section{Introduction}\label{sec:intro}

Cosmic structure formation is hierarchical \citep{white_rees_78}, and galaxy mergers are expected to result in gravitationally bound massive black hole binaries (MBHBs) with masses in the range $10^{5-10} {\rm M}_{\astrosun}$ \citep{begelman_80}. Stellar interactions and gas torques can shrink the binary's orbit and cause the binary to enter the gravitational wave (GW) driven inspiral regime. The upcoming Laser Interferometer Space Antenna (LISA, \citealt{lisa_paper, lisa_report_new}) and pulsar timing array (PTA) experiments are expected to be able to observe the GW emission from the mergers of these MBHBs. The cumulative emission from the cosmological population of supermassive black hole binaries provides the most natural explanation for the recently discovered stochastic GW background \citep{nanograv_gw_bg}.

GW emission alone will provide an unequivocal first detection of an MBHB merger, as well as deliver rich new tests of general relativity (GR) and other science. However, when combined with electromagnetic (EM) observations, the GW data will open new windows on additional astrophysics and fundamental physics \citep{baker_haiman_multimessenger_review, bogdanovic_em_bh_review}. For example, with both GW and EM signals we can better constrain the properties---such as eccentricity, semi-major axis, and mass---of MBHBs, probe novel regimes of accretion physics, test the co-evolution of massive black holes and their host galaxies, as well as improve our understanding of key fundamental physics by probing alternative gravity theories \citep{Hazboun_2013,deRham_2018}, constraining the nature of inflation and dark matter \citep{Caprini_2018}, and measuring the redshift-distance relation \citep{Schutz_86} as well as the mass of the graviton \citep{kocsis_08, hassan_2012,kelley19, bogdanovic_em_bh_review}.

MBHBs accreting from circumbinary gas have been found to exhibit three sources of unique, periodic, electromagnetic emission: relativistic Doppler modulation of the apparent brightness of the black holes' (BHs) accretion disks \citep{dorazio_spikey,hu_spikey,charisi_doppler}, gravitational self-lensing \citep{dorazio2018,ingram2021,davelaar2022,davelaar2022b,krauth2023}, and periodicities induced by the system's hydrodynamical variability \citep{bode_circular_lc, giacomazzo_lc_circle,ryan_sailfish,luke_decoupling}. In this paper, we will focus on the latter source of periodicity.

The hydrodynamic behavior of equal-mass, circular binaries embedded in circumbinary disks (CBDs) has been well studied \citep{bode_circular_lc,giacomazzo_lc_circle, gutierrez_circle_lc, Paschalidis_lc_circular, farris_2014, farris_2015, farris_2015_2, sb_code_comp}. An important recent finding is that torques from the circumbinary gas disk can induce a significant equilibrium eccentricity in the binary's orbit, in the range of $e \approx 0.4-0.5$ \citep{munoz_miranda_lai_19, zrake_eqbm_ecc, duffell_dorazio_2020, siwek_orbevol}. Several hydrodynamical simulations have also focused on the reverse question and studied the impact of eccentric binaries on the hydrodynamics of circumbinary disks. They have detailed the disk's morphology and behavior \citep{miranda_munoz_lai_17,siwek_orbevol}, accretion rate \citep{munoz_lai_16, siwek_pref_accretion}, as well as light-curves \citep{Bogdanovic_08_lc, ryan_sailfish}.

Recently, hydrodynamical simulations have also begun to study the apsidal precession of binaries. Indeed, non-zero rates of disk-induced binary precession have been reported across a range of masses \citep{tiede_disk_prec,dittmann_prec,calcino_prec}. In addition, compact binaries that are eccentric and approaching relativistic orbital speeds will also experience apsidal precession due to general relativity. This additional apsidal precession of binaries is expected to impact the hydrodynamics of the circumbinary gas further, but its effects have not been explored to date. 

In this paper, we present results from hydrodynamical simulations of equal-mass, eccentric, massive black-hole binaries that experience relativistic precession. Our goal is not only to assess key hydrodynamic features of generic, apsidally precessing binaries, general-relativistic or not, in circumbinary disks but also to understand the impacts of binary precession on pre-merger EM signals, constructing a complete picture of the multi-messenger behavior of MBHB mergers. 

Our paper is organized as follows.
In \autoref{sec:methods}, we discuss our governing equations of motion, hydrodynamical code, post-processing spectral analysis, simulation setup, and numerical methods. 
In \autoref{sec:morphology}, we discuss the morphology of gas found in our simulations, mainly focusing on the central cavity and its evolution. 
In \autoref{sec:strong_gr}, we detail the simulation results of $e=0.45$ binaries. We discuss the systems' accretion rate (\autoref{ssec:accretion_rate}) and expected light-curves (\autoref{ssec:light-curves}), highlighting the signal's periodicity (\autoref{sssec:large_scale_signals}) and the mechanisms behind the observed flares (\autoref{sssec:flaring}).
In \autoref{sec:weak_gr}, we discuss the results of our $e=0.15$ binary simulations, detailing how the cavity evolves and how the binary accretes from it (\autoref{ssec:damped_pref}, \autoref{ssec:cavity_precession}, \autoref{ssec:circ_binary_accretion}, \autoref{ssec:circ_accretion_ratio}), as well as the periodicity of its accretion and EM light-curves (\autoref{ssec:beat}). 
In \autoref{sec:observations}, we discuss practical constraints to ensure that the EM signal is observable (\autoref{ssec:cavity_req}, \autoref{ssec:timescales}) and that the system exhibits an EM signal while it is simultaneously detectable by {\em LISA} (\autoref{ssec:gw_detection}). 
Finally, we summarize our main conclusions and the implications of this work in \autoref{sec:conclusions}.

\section{Methods}\label{sec:methods}

\subsection{Equations of motion}\label{ssec:eom}

We model the evolution of our binaries according to Kepler's laws of motion with post-Newtonian corrections. The solution to Kepler's equations in polar coordinates is given by
\begin{equation}
    r = \frac{a \left( 1-e^{2} \right)}{1+e\left( \cos \left( v \right) \right)},
\end{equation}
where $r$ is the radius, $a$ is the semi-major axis, $e$ is the eccentricity, and $v \equiv \phi - \varpi$ is the true anomaly, defined as the angular position of the body ($\phi$) relative to the longitude of its pericenter ($\varpi$). The solution to Kepler's equations as a function of time, however, we obtain numerically. We note that

\begin{equation}
    r=a \left( 1-e\cos \left(u \right) \right)
\end{equation}
\begin{equation}\label{eqn:root_kep}
    l \equiv n (t-t_{0}) = u-e\sin \left( u \right)
\end{equation}
where $l$ is the mean anomaly, $n$ is the mean motion, and $u$ is the eccentric anomaly. We calculate $u$ at each time step by numerically finding the root of \autoref{eqn:root_kep}, keeping relative errors under $10^{-12}$. 

General relativity can be approximated as deviations from Newton's laws via a power series in $\rm{v}/c$, where $\rm{v}$ is a characteristic speed in the system. The first two post-Newtonian (PN) orders (i.e.~up to $(\rm{v}/c)^4$) conserve the energy of the system and describe relativistic precession, whereas dissipative effects enter at 2.5th PN order (i.e.~$(\rm{v}/c)^5$), representing gravitational wave radiation that shrinks the binary's semi-major axis and eccentricity. Since, in this study, we are interested in the effects of sustained precession and not the orbital decay due to GW radiation, we consider relatively wide binaries, for which the orbital decay due to the 2.5PN term is negligible during our simulation runs.

The 2PN Quasi-Keplerian (QK) equations of motion can be written as

\begin{equation}
    r = a_{r} \left( 1-e_{r}\cos \left( u \right) \right)
\end{equation}
\begin{equation}
    v=2\arctan\left[\left(\frac{1+e_{\phi}}{1-e_{\phi}}\right)^{1/2}\tan\left(\frac{u}{2}\right)\right]
\end{equation}
\begin{equation}\label{eqn:root_gr}
\begin{aligned}
    l \equiv  n \left( t-t_{0} \right) = & \ u - e_{t}\sin \left( u \right) + \\
    & \frac{g_{4t}}{c^{4}} \left( v-u \right) + \frac{f_{4t}}{c^4}\sin \left( v \right)
\end{aligned}
\end{equation}
\begin{equation}\label{eqn:phi_def}
    \frac{2\pi}{\Phi}(\phi) = v + \frac{f_{4\phi}}{c^4}\sin(2v) + \frac{g_{4\phi}}{c^{4}}\sin(3v)
\end{equation}
\begin{equation}
\begin{aligned}
   n = &  {} (-2E)^{3/2}(1 + \frac{-2E}{8c^{2}}(-15 + \eta) + \\
    & \frac{(-2E)^{2}}{128 c^{4}}(555 + 30\eta + 11\eta^{2} + \\
    & \frac{192}{\sqrt{-2Eh^2}}(-5 + 2\eta) ) ),
\end{aligned}
\end{equation}
where $E$ and $h$ represent the reduced energy and reduced angular momentum of the binary, respectively. $\eta \equiv \mu/M$ is the finite mass ratio, $v$ is the true anomaly, and $\Phi$ represents the angle subtended by the periastron per orbital revolution. The quantities with subscripts are gauge-dependent and uniquely defined for given values of $E$, $h$, and $\eta$, and
are derived in full in Section 10 of \citealt{blanchet_review}. 
To solve this set of equations in time for a given system, we need only define $E$, $h$, and $\eta$,  which can be calculated directly from the binary's eccentricity ($e \equiv e_{r}$), masses $(M_1, M_2)$ and semi-major axis ($a \equiv a_{r}$). Just as for the Newtonian case, we solve \autoref{eqn:root_gr} at each time-step via Newton-Raphson root-finding, keeping relative errors under $10^{-12}$.

\subsection{Hydrodynamical simulations}\label{ssec:hydro}

All simulations were run using the publicly available GPU-accelerated finite-volume grid code \verb|Sailfish|. Full technical descriptions of the code can be found in \citealt{ryan_sailfish}; here, we provide a brief description of its main relevant features.

\verb|Sailfish| is a Cartesian grid-based code that solves the following vertically-integrated Newtonian fluid equations under the assumption of a thin-disk ($h/r \ll 1$) and mirror symmetry about the midplane ($z=0$):
\begin{equation}
    \partial_{t}\Sigma + \nabla_j (\Sigma v^j) = S_\Sigma
\end{equation}
\begin{equation}
    \partial_t  (\Sigma v_i) + \nabla_j(v^j v_i + \delta_{i}^{j}\mathcal{P}) = g_{i} + \nabla_j{\tau_{\rm{visc}}}_i^j + S_{p,i}
\end{equation}
\begin{equation}
    \partial_t E + \nabla_j [(E + \mathcal{P})v^j] = v^jg_j + \nabla_j (v^i {\tau_{\rm{visc}}}_i^j) - \dot{Q} + S_E.
\end{equation}
Here $\Sigma$ is the surface density of the disk, $\mathcal{P}$ is the vertically integrated pressure, $v^i$ is the mid-plane horizontal fluid velocity, $E$ is the vertically integrated energy density, $g^{i}$ is the vertically integrated gravitational force density, 
\begin{equation}
    {\tau_{\rm{visc}}}_i^{j}  = \Sigma \nu \left(  \nabla_i v^j + \nabla^j v_i - \left(2/3\right) \delta^j _i \nabla_k v^k \right)
\end{equation}
is the viscous stress-tensor, and $S_{E}$, $S_{\Sigma}$, $S_{p,i}$ are the energy, mass, and momentum\footnote{For simplicity we use a torque-free sink model, but see \citealt{dittmann_torque_free} for other torque-controlled options.} sink terms. $\dot{Q}$ represents the cooling of hydrogen-dominated gas via vertical transport of heat by radiative diffusion \citep{Frank_2002}, given by
\begin{equation}
    \dot{Q} = \frac{8}{3} \frac{\sigma}{\kappa \Sigma} \left(\frac{m_{p}\mathcal{P}}{\kappa_{B}\Sigma}\right)^4,
\end{equation}
where  $\sigma$ is the Steffan-Boltzman constant, $\kappa = 0.4~ \rm{g}\, \rm{cm}^{-2}$ is the electron scattering opacity, $m_p$ is the proton mass, and $\kappa_{B}$ is the Boltzmann constant.

We embed a binary in a Shakura-Sunayev $\alpha$-disk \citep{shakura_sunayev_1973} with viscosity parameter $\alpha=0.1$, and prescribe a gamma-law equation of state $\mathcal{P}=\Sigma \epsilon(\Gamma -1)$ with $\Gamma=5/3$, where $\epsilon$ is the mid-plane specific internal energy. Though radiation pressure is often important in black hole accretion disks, it is numerically challenging to include it in simulations. Rather, we follow \citet{ryan_sailfish} by omitting radiation pressure from the implementation while selecting characteristic disk aspect ratios in the initial conditions that are representative of disks whose gas and radiation pressures are accounted for. In practice, we select aspect ratios in these ``gas pressure-only models'' by specifying artificially high accretion rates in the initial conditions. A gas-pressure only model thereby approximates its gas- and radiation-pressure ``target model,'' and an adjustment is applied in post-processing to further facilitate this approximation --- see \autoref{ssec:post_processing}. We relate the effective temperature of the disk to the mid-plane temperature via 
\begin{equation}
    T_{\rm{eff}}^4 = \frac{4}{3} \frac{T^4}{\kappa \Sigma},
\end{equation}
which is valid for a non-trivial mixture of gas and radiation pressure \citep{ryan_sailfish}.

\subsection{Post-processing spectral analysis}\label{ssec:post_processing}

We obtain the effective temperature from
\begin{equation}
    \dot{Q}= 2 \sigma T_{\rm{eff}}^4,
\end{equation}
where the factor of 2 represents the 2 radiating faces of the disk. Since $T_{\rm{eff}}^4$ scales linearly with $\dot{M}$, we first scale $T_{\rm{eff}}^{4}$ down to the target model's accretion rate of $10\dot{M}_{\rm{Edd}}$ that includes both gas and radiation pressure (see \citealt{ryan_sailfish} for more details of the scaling procedure). Then, we compute a ``multi-color blackbody'' spectrum by assuming that each cell radiates as a black body set by its effective temperature and obtain the luminosity in discrete frequency bands by integrating the surface luminosity density
\begin{equation}
    \frac{dL}{da} =  \int_{\rm{f}_{1}}^{\rm{f}_{2}} \frac{2h \rm{f}^3 / c^2}{\exp(\frac{h\rm{f}}{\kappa T_{\rm{eff}}})-1} \,d\rm{f}
\end{equation}
over the simulation domain, where $\rm{f}$ is the photon's frequency and $da$ is the cell area. Notably, we do not account for Doppler brightening effects, and thus, our light-curves are valid for systems viewed sufficiently far from edge-on. Finally, we define the X-ray, UV, and optical bands to correspond to the wavelength ranges of $[10^{-10},10^{-7}]$ $[10^{-7},4 \times 10^{-5}]$ $[4 \times 10^{-5}, 7 \times 10^{-5}]$ cm, respectively.

\subsection{Simulation setup}\label{ssec:design}

\begin{table}
\centering
\caption{Parameters of our simulated binaries.}

\resizebox{\columnwidth}{!}{

\begin{tabular}{|c|c|c|c|}
 \hline
 Eccentricity& Semi-Major Axis& Precession Rate&  Precession Period\\
 $e$& $a$ [$R_{S}$]& $\Phi - 2\pi$ [$\rm{deg}/\tau_{\rm{GR}}$]&  $\tau_{\rm{Prec}}$ [$\tau_{\rm{Kep}}$]\\
 \hline
$0.45$   & $150$ & $4.65$ & $38.96$ \\
$0.45$   & $175$ & $3.97$ & $45.61$ \\
$0.15$   & $175$ & $3.77$ & $48.04$ \\
\hline
\end{tabular}

}
\label{tab:hires_params}
\tablecomments{Here, $R_{S}\equiv 2G(M_{1}+M_{2})c^{-2}$, $\Phi$ is calculated from \autoref{eqn:phi_def}, $\tau_{\rm{Kep}}$ and $\tau_{\rm{GR}}$ are the orbital periods for the Keplerian and GR binaries, respectively, and $\tau_{\rm{Prec}}$ is the time, in terms of Keplerian orbits, for the binary's longitude of periapsis to precess by $\pi$.}

\end{table}

We investigate the nature and impact of general relativistic precession of MBHBs in hydrodynamic simulations by running ``twin'' pairs of nearly-identical hydrodynamic simulations which only differ in the binary's equation of motion---Keplerian or Quasi-Keplerian. We then compare the results to understand how the inclusion of precession affects the system. However, in order to gain broader insight into the nature of binary precession, we run a suite of simulations that sample different binary eccentricities, semi-major axes, and, as a byproduct, a range of precession rates. Namely, we consider binaries with eccentricities of $e=0.45$ and $e=0.15$ and separations of $150~R_S$ or $175~R_S$ (where $R_{S}\equiv 2~G(M_{\rm bin})/c^2$ is the Schwarzschild radius for the total binary mass $M_{\rm bin}$). Our choice of $e=0.45$ is motivated by the equilibrium value found in \citealt{zrake_eqbm_ecc}. We list our simulation parameters in \autoref{tab:hires_params}.

We limit our simulations in this study to component masses $M_1$, $M_2$ satisfying $M_2/M_1 \equiv q=1$ with total mass $M_{\rm{bin}} \equiv M_{1} + {M_2}=10^{7}~M_{\astrosun}$, and an orbital mach number $\mathcal{M}=(1/ \sqrt{\Gamma}) (h/r)^{-1} \simeq 11$. We denote the Keplerian binary orbital period as $\tau_{\rm Kep}$.

\subsection{Numerical choices}\label{ssec:numerics}

Our simulations are run on a square Cartesian grid with side length $2D$, where $D \equiv 15a$ is the domain radius. The sink radii are set to be $0.05a$ and the gravitational force derives from a Plummer potential with softening length $r_s$ set equal to the sink radius (see \citet{ryan_sailfish} for further details). Further, we initialize the disk with the following near-equilibrium profiles, after which the disk dynamically relaxes:

\begin{equation}
    \Sigma = \Sigma_0 \left( \frac{r_{\rm{soft}}}{a}  \right)^{-3/5}
\end{equation}

\begin{equation}
    \mathcal{P} = \mathcal{P}_0 \left( \frac{r_{\rm{soft}}}{a}  \right)^{-3/2}
\end{equation}

\begin{equation}
    v_{\theta} = \sqrt{\frac{GM_{\rm{bin}}}{r_{\rm{soft}}}}
\end{equation}
where $r_{\rm{soft}}=\sqrt{r^2 + r_s^2}$ and $\Sigma_0 \approx 0.03 ~M_{\rm{bin}}/a^2$ and $\mathcal{P}_0 \approx 10^{-4} ~M_{\rm{bin}}\Omega_{\rm{bin}}^2$. These profiles imply a temperature profile of 
\begin{equation}
    T = \frac{m_p \mathcal{P}_0}{k_B \Sigma_0}  \left( \frac{r_{\rm{soft}}}{a}  \right)^{-9/10}
\end{equation}

We define the viscous inflow rate $\dot{M}_{\rm{vis}}\equiv 3\pi\Sigma_0 \nu_{\rm{vis}}$, where $\nu_{\rm{vis}}$ is the kinematic shear viscosity. The binary is initialized at its pericenter, with the semi-major axis being along the x-axis. 

We employ ``buffer'' source terms from $r=D-0.1a$ to $r=D$, that drive the system to the initial disk conditions and prevent boundary effects from propagating into the interior of our disk. We exclude the buffer region from any of our post-processing calculations described in \autoref{ssec:post_processing}. Further details on the buffer region can be found in Section 3 of \citealt{ryan_sailfish}.

Our idealized, axisymmetric initial conditions do not accurately reflect the viscous disk around a binary. Thus, we relax the disk into a quasi-steady state by first evolving the system over three characteristic viscous times, before data is taken. To do this at minimal computational cost, we employ a method of ``up-sampling'', where the resolution of the simulation is gradually increased throughout the run. Namely, we run a $dx=0.03a$ resolution $10^{6}$ cell grid for $720~ \tau_{\rm{Kep}}$, followed by a $dx=0.015a$ $4 \times 10^{6}$ cell grid for $180~ \tau_{\rm{Kep}}$, before up-sampling to a $16 \times 10^{6}$ cell grid --- a final resolution of $dx=0.0075a$. We then begin recording data and continue to evolve the simulation for $100~ \tau_{\rm{Kep}}$. Simulation snapshots were output at a cadence of $0.1~\tau_{\rm{Kep}}$ from $t=900$-$950~ \tau_{\rm{Kep}}$, and then at a cadence of $0.01~ \tau_{\rm{Kep}}$ until t=$1000~\tau_{\rm{Kep}}$.\footnote{This was done to study the system at a high time-resolution with minimal storage cost; each $16 \times 10^{6}$ cell grid snapshot has a file-size of approximately $0.5\,$GB.}

All our simulations were run on Columbia University’s High-Performance Computing Cluster \textit{Ginsburg} using NVIDIA A100 GPUs. Each $16 \times 10^{6}$ cell grid completed $\approx 0.36 ~\tau_{\rm{Kep}}$ per GPU hour. Employing our up-sampling method, the total run-time for each simulation from $t=0$ to $t=1000 ~\tau_{\rm{Kep}}$ was $\approx 1.7 \times 10^{5}$ GPU hours.

Finally, we note that our simulations are explicitly 2D. Thus, we do not account for the BHs' spin nor nodal GR precession. Furthermore, our binary's orbital motion is fixed and prescribed by the equations defined in \autoref{ssec:eom}. While we do not model the binary's inspiral, we note that the binary shrinks minimally over the time-scale on which we record data---shrinking by $\Delta a / a< 1 \%$ over $100 ~\tau_{\rm{Kep}}$.

\section{System morphology}\label{sec:morphology}
\begin{figure}
    \centering
    \includegraphics[width=1\columnwidth]{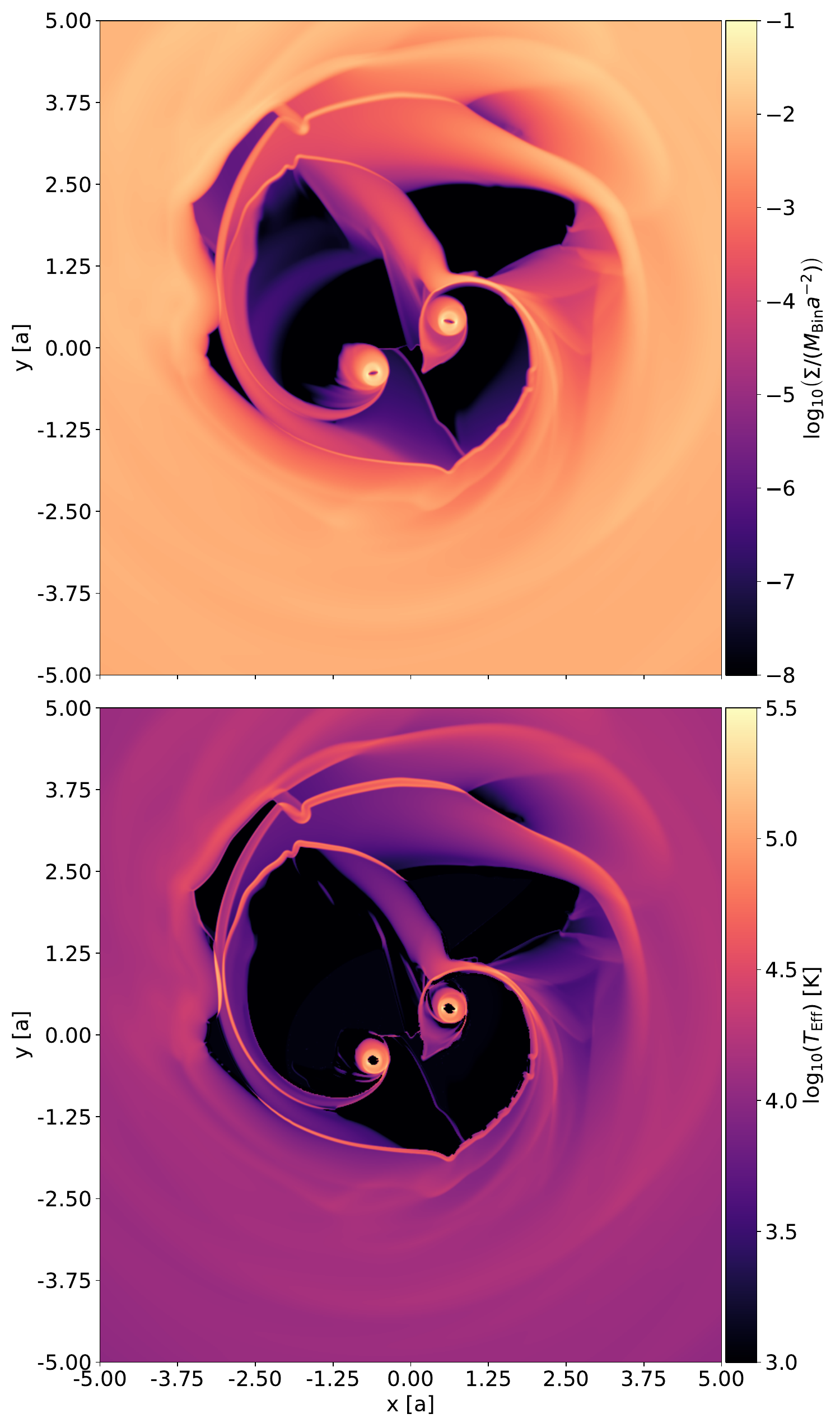}
    \caption{A representative snapshot of a hydrodynamical simulation of MBHBs in CBDs. This snapshot is from the $e=0.45$, $a=150~R_S$, GR simulation. The upper and lower panels display the surface density and effective temperature maps.}
    \label{fig:sample_snap}
\end{figure}

We begin by briefly recapitulating the most striking features present in a hydrodynamic simulation of a MBHB embedded in a CBD. Similar features have been found in many previous studies. \autoref{fig:sample_snap} displays representative snapshots of the surface density and effective temperature distributions for the $e=0.45$, $a=150~R_S$, GR case.

We note the presence of an under-dense central region, often referred to as a central cavity. The sharp outer edge of this region is the cavity wall. Inside the cavity, there are accretion disks surrounding each of the individual BHs, called ``mini-disks." Lastly, note the central cavity is not entirely evacuated but is instead filled with distinct, extended filaments of gas. These filaments, or ``streams," are generated by the binary's gravitational influence, which tends to peel gas from the cavity wall and produce these streams~\citep{Dorazio+2013}.

Finally, we note that the hottest regions in our simulation domain are the mini-disks, with effective temperatures of $T_{\rm{eff}} >10^{5} ~K$. However, we also note that regions where the streams collide with each other or with the cavity wall are hot as well, with $T_{\rm{eff}} >10^{4.5}~K$. The rest of the gas in our simulation remains at a relatively uniform $T_{\rm{eff}}\approx 10^{4.25}~K$.

\begin{figure}
    \centering
    \includegraphics[width=1\columnwidth]{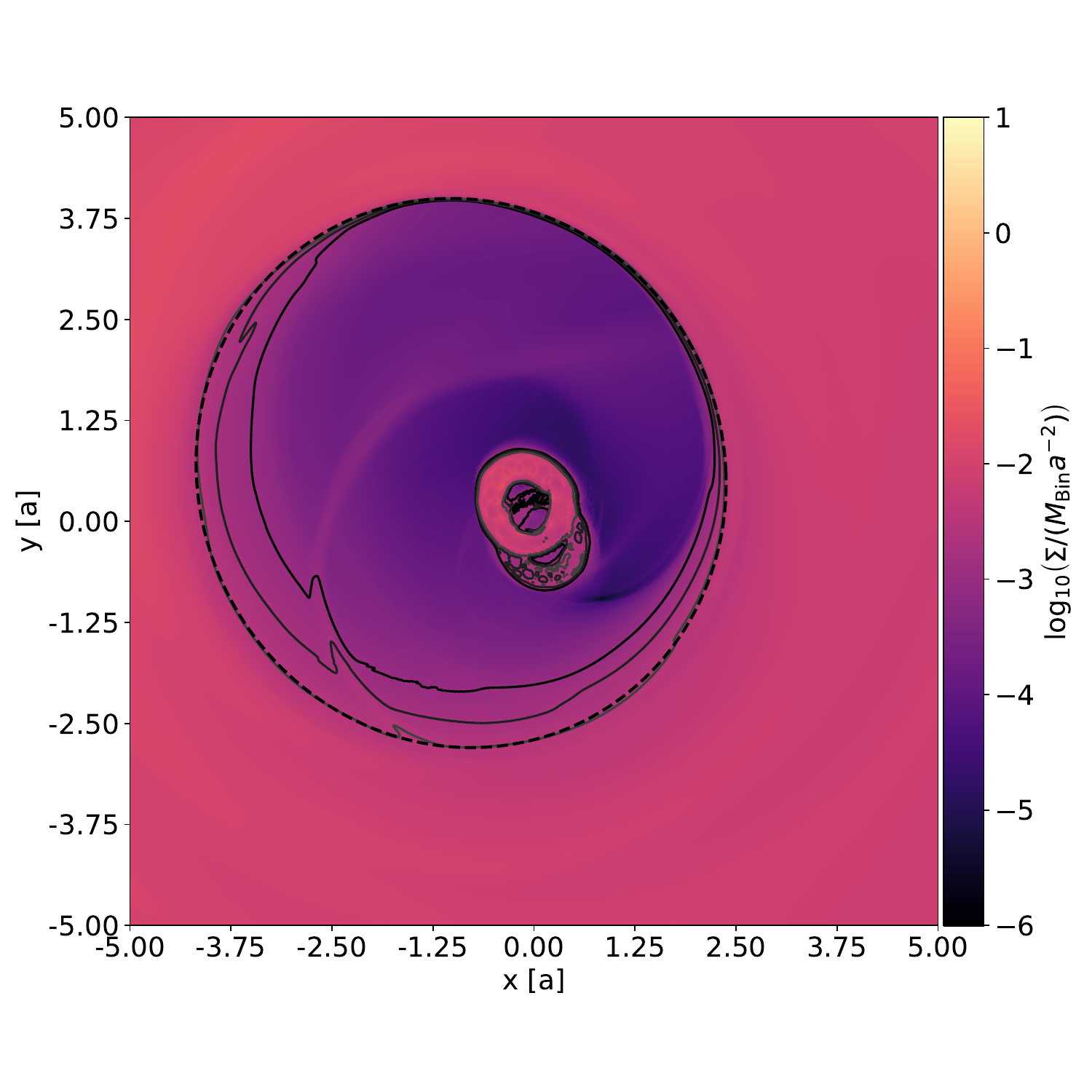}
    \caption{Surface-density map for the $e=0.45$, $a=175~R_S$, GR simulation, averaged over one orbit. The solid lines represent surface-density contours of $\Sigma=10^{-3}$, $10^{-2.75}$, and $10^{-2.5} ~M_{\rm{bin}}/a^{2}$, with the lightest line representing the $10^{-2.5} ~M_{\rm{bin}}/a^{2}$ contour. The black dotted line shows the best-fit ellipse to the $10^{-2.5} ~M_{\rm{bin}}/a^{2}$ contour. The two small elliptic rings within the cavity are the mini-disks completing a full orbit.}
    \label{fig:cavity}
\end{figure}

\subsection{Cavity features}

\begin{table*}
    \centering
    \caption{Parameter values of ellipse fit to time-averaged cavities.}
\resizebox{\textwidth}{!}{
\begin{tabular}{|c|c|c|c|c|c|c|c|c|c|c|}
 \hline
 $e$& $a$ $[R_S]$& GR Precession& ($x_{\rm{C}}$, $y_{\rm{C}}$)& ($a_{\rm{C}}$, $b_{\rm{C}}$)& $\theta_{\rm{C}}$& $r_{\rm{C}}$& ($\Delta x_{a}$, $\Delta y_{b}$)\\
 \hline
$0.45$& $175$& Yes& $(-0.85,0.58)$& $(3.45,3.32)$& $2.11$&  $1.03$& $(0.93,0.43)$\\
$0.45$& $175$& No&  $(0.66,1.12)$& $(3.69,3.42)$& $0.88$&  $1.30$& $(1.29, 0.21)$\\
$0.45$& $150$& Yes& $(1.03,-0.26)$& $(3.66,3.44)$& $2.59$&  $1.06$& $(1.01,0.31)$\\
$0.45$& $150$& No&  $(0.02,-0.94)$& $(3.49,3.33)$& $1.40$&  $0.94$& $(0.92,0.18)$\\
$0.15$& $150$& Yes& $(0.30,0.07)$& $(2.13,2.02)$& $2.05$&  $0.30$& $(0.07,0.29)$\\
$0.15$& $150$& No&  $(-0.17,-0.16)$& $(1.98,1.90)$& $2.26$& $0.23$& $(0.01,0.23)$\\

\hline

\end{tabular}

}

\label{tab:cavity_params}
\tablecomments{Parameter values of ellipses fit to the $\Sigma=10^{-2.5}~M_{\rm{bin}}/a^2$ and $\Sigma=10^{-3} ~M_{\rm{bin}}/a^2$ surface-density contour for $e=0.45$ and $e=0.15$ simulations, respectively, as a proxy for the size, shape and location of the central cavity. Listed are the median values over $100$ fits, with each ellipse fitted to the time-averaged surface density map over a single full orbit.  $x_{\rm{C}}$, $y_{\rm{C}}$ are the coordinates of the center of the ellipse, $a_{\rm{C}}$, $b_{\rm{C}}$ are the semi-major and semi-minor axes, $\theta_{\rm{C}}$ is the azimuthal direction of the cavity's semi-major axis, $r_{\rm{C}}$ is the distance between the center of the cavity and the barycenter, and $\Delta x_{a}$ and $\Delta y_{b}$ represent the shortest distance from the barycenter to the cavity's semi-major and semi-minor axes, respectively. Our coordinate system's origin is the binary barycenter. Length units are the binary semi-major axis $a$, and angles are reported in radians from the positive x-axis.}
\end{table*}

Note how the propagating tidal streams in \autoref{fig:sample_snap} make it difficult to visualize the cavity. However, by time-averaging the surface density over a binary orbit, we reveal the cavity's average properties more clearly. We describe the time-averaged cavity geometrically by fitting an ellipse to the $\Sigma=10^{-2.5} ~M_{\rm{bin}}/a^{2}$ and $\Sigma=10^{-3} ~M_{\rm{bin}}/a^{2}$ surface-density contour for $e=0.45$ and $e=0.15$ simulations, respectively. These density contours provide good qualitative representations of the cavity in each case -- see the example in \autoref{fig:cavity}. In \autoref{tab:cavity_params}, we detail the median parameter values of these ellipses. 

From \autoref{tab:cavity_params} and \autoref{fig:cavity}, it is clear that the cavity properties are unique to each simulation, changing with the binary's eccentricity, semi-major axis,\footnote{Although length scales in our simulations are normalized to the initial binary semi-major axis, the system is not scale-free with respect to it because the radiative cooling term described in \autoref{ssec:post_processing} breaks scale symmetry.} and equations of motion. A general understanding of how the cavity's geometrical shape responds to different initial binary conditions is beyond the scope of this paper. Nevertheless, we describe the correlations we find in our simulations. 

We first note that cavity size shows some correlation with binary eccentricity, with the $e=0.45$ simulations having larger cavity semi-major and semi-minor axes than the $e=0.15$ simulations. We also find that the $e=0.15$ simulations are more symmetric, in the sense that the center of the cavity is less offset from the binary's barycenter, when compared to the $e=0.45$ simulations. We also note that none of the cavities are centered about the binary barycenter, and their semi-major axes are not oriented along the binary's semi-major axis.

\subsection{Cavity evolution}
\begin{figure}
    \centering
    \includegraphics[width=\columnwidth]{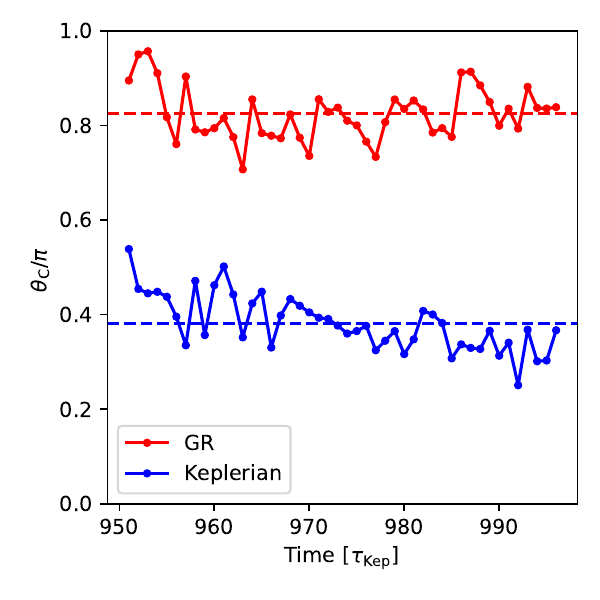}
    \caption{Time evolution of the orientation of the cavity's semi-major axis, $\theta_{\rm{C}} \ [\rm{radians}]$, as measured from the barycenter in the $e=0.45$, $a=150~R_S$ simulation. The data for the single-orbit-averaged GR and Keplerian cavities are shown in blue and red, respectively. The median values are shown with similarly colored, dashed horizontal lines. We note that the orientation of the GR simulation's cavity is relatively constant over the precession period $\tau_{\rm{Prec}}=38.96~\tau_{\rm{Kep}}$.}
    \label{fig:phi_cavity}
\end{figure}

\begin{figure}
    \centering
    \includegraphics[width=1\columnwidth]{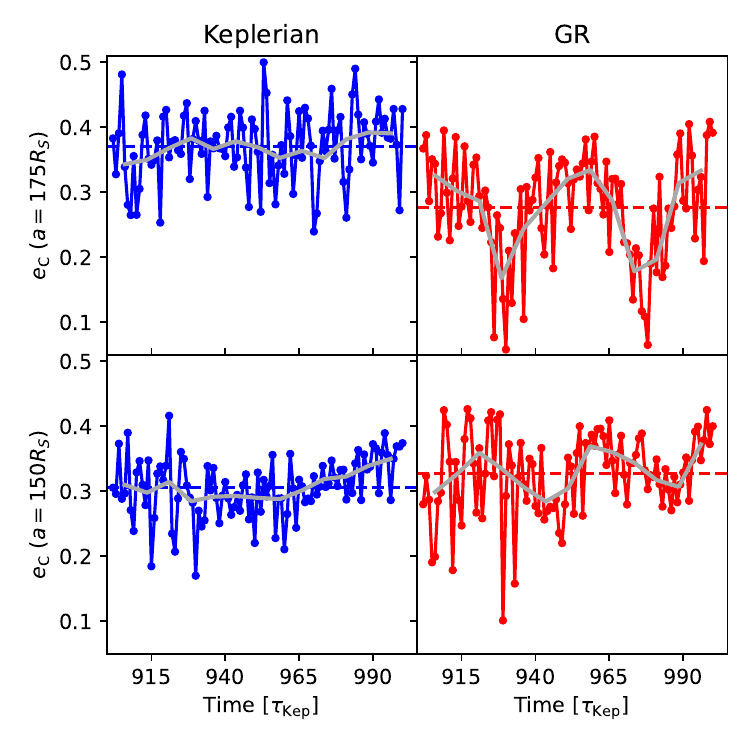}
    \caption{The $e=0.45$ cavities' eccentricity ($e_{\rm{C}}$) evolution. The upper and lower panels display the results for the $a=175~R_S$ and $a=150~R_S$ cases, respectively. The Keplerian case is displayed on the left in blue, and the GR case on the right in red. The moving median with a window of $7~\tau_{\rm{Kep}}$ is overlaid in gray in each panel. The total median eccentricity value is displayed as a horizontal dashed line.}
    \label{fig:ecc_time_cavity}
\end{figure}
Next, we describe how the orbit-averaged cavity features of the $e=0.45$ simulations evolve over time---the features of the $e=0.15$ cavities are detailed in \autoref{sec:weak_gr}.

First, we note that the cavity semi-major axes in the $e=0.45$ runs are oriented in a constant direction. This is seen in both the GR and Keplerian runs and is depicted well in \autoref{fig:phi_cavity}. Note how the orientation of the Keplerian and GR cavity, though noisy, both remain close to their median values, even though the binary semi-major axis of the GR simulation has precessed by an angle $\pi$ over a time we define to be $\tau_{\rm{Prec}}$ (see \autoref{tab:hires_params} for values).

Although the orientation of the cavity's semi-major axis ($\theta_{\rm{C}}$) remains constant over time, the eccentricity of the cavity ($e_{\rm{C}}$) does not, and appears to depend on the precession period. In \autoref{fig:ecc_time_cavity}, we display the eccentricity evolution of our $e=0.45$ simulations' cavities. While all our simulations display some evolution, the GR simulations' cavity eccentricities exhibit periodic oscillations on timescales of tens of binary orbits, whereas the Keplerian simulations' are more or less constant, with no coherent trends. Further, we note that the amplitude of the periodicity is larger in the $a=175~R_S$ case (top right panel), which is the binary with slower GR precession. Finally, in both cases, we note that the period of this eccentricity oscillation agrees well with the precession periods of $45.61 ~\tau_{\rm{Kep}}$ and $38.96 ~\tau_{\rm{Kep}}$, i.e. we measure peak periods of $45 ~\tau_{\rm{Kep}}$ and $38 ~\tau_{\rm{Kep}}$ for the $a=175~R_S$ and $a=150~R_S$ simulations, respectively. 

Since the eccentricity-oscillation occurs only in the GR simulations, and has a period $\approx \tau_{\rm{Prec}}$, it is clearly caused by the precession of the binary. However, the strength of the effect being anti-correlated with the precession rate is not as apparent. We suggest that the eccentricity modulation is caused by a similar mechanism that is responsible for the eccentricity growth of the Keplerian cavity. Hydraulic pumping (see \citealt{shi_2012,lai_munoz_review}) and eccentric Lindblad resonances (see  \citealt{ hirose_osaki_90, lubow_91a, lubow_91b,lai_munoz_review}) are both plausible causes, but the hydraulic pumping mechanism is particularly appealing due to the amplitude's anti-correlation with precession rate.

For slower precession, the binary's apocenter shifts less from one orbit to the next. Thus, consecutive streams are launched toward the cavity in a more coherent direction. Stream impacts on the cavity are more angularly coherent, pushing the impacted area of the cavity wall back further and driving stronger eccentricity modulations. 

While further analysis of this mechanism remains outside the scope of this paper, in future work, we plan to investigate the effect of binary precession on the morphology and evolution of the cavity and, more broadly, of the circumbinary disk, especially since these could affect gas-torques on the binary, and may have significant effects on disk-induced binary orbital evolution.

\section{Eccentric binaries}\label{sec:strong_gr}

In the following section, we detail the key characteristics of our eccentric ($e=0.45$) binary simulations. In \autoref{ssec:accretion_rate}, we discuss the accretion rate of the system, demonstrating how its modulation is uniquely attributable to precession. In \autoref{ssec:light-curves}, we present and analyze the expected light-curves of the system, characterizing and determining the mechanism for both the long time-scale (tens of orbits) signal (\autoref{sssec:large_scale_signals}) as well as the multiple individual flares (or sub-peak structures) that occur each orbit (\autoref{sssec:flaring}). Throughout this section, we will conduct our analysis considering an MBHB in a static cavity. While this is not precisely the case, it is a helpful approximation for our $e=0.45$ simulations. 

\subsection{Accretion rate}\label{ssec:accretion_rate}

\begin{figure}
    \centering
    \includegraphics[width=1\columnwidth]{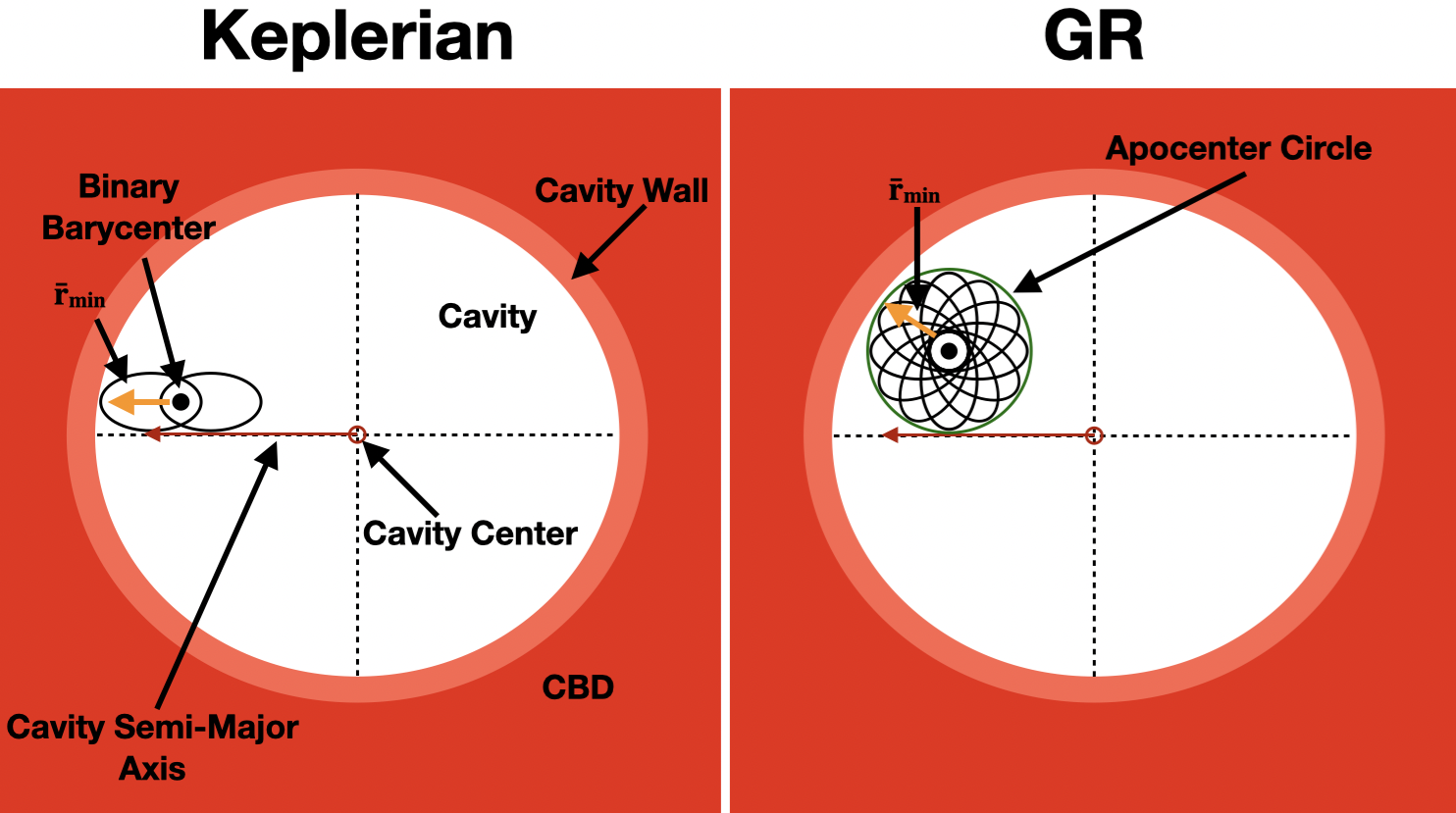}

    \caption{Diagram of the binary-cavity geometry. The left and right panels display the Keplerian and GR binaries, respectively. $\vec{r}_{\rm{min}}$ is the vector that points from the binary barycenter to the closest part of the cavity wall and thus defines the azimuthal position at which the ratio of the two BHs' accretion rates is at an extremum. If the binary's semi-major axis is aligned with  $\vec{r}_{\rm{min}}$, one BH will spend more time closer to the cavity than the other, thereby accreting more. As depicted, the binary apsides in the Keplerian case have a fixed orientation, so the same BH will be closer to the cavity, resulting in a stable accretion ratio over time. The semi-major axis of the GR binary, however, precesses. This precession modulates which BH is closer to the cavity and generates a quasi-sinusoidal accretion ratio. Finally, we note how $\vec{r}_{\rm{min}}$ is dependent on the position of the barycenter.
    \label{fig:diagram}
    }
\end{figure}

\begin{figure}
    \centering
    \includegraphics[width=1\columnwidth]{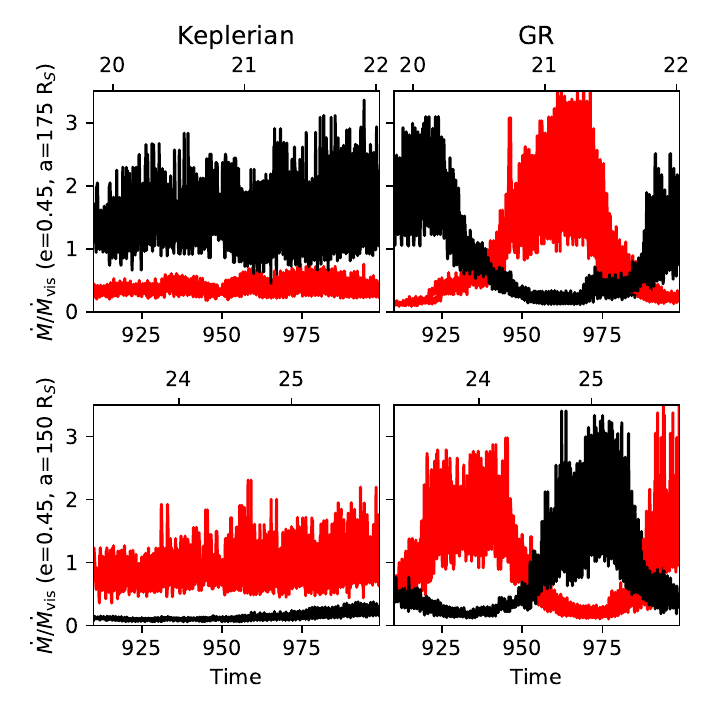}
    \caption{
    The $e=0.45$ simulations' accretion rates over time. The upper and lower rows display the $a=175~R_S$ and $a=150~R_S$ simulations, respectively. The left panels show the Keplerian simulations, which exhibit constant preferential accretion, while the right panels display GR simulations, which exhibit quasi-sinusoidal accretion rate modulations. The accretion rates are normalized by the viscous inflow rate $\dot{M}_{\rm{vis}} \equiv 3\pi\Sigma_0\nu_{\rm{vis}}$, with $\dot{M}_{1}$ plotted in red and $\dot{M}_{2}$ plotted in black. Time is given on the horizontal axes in different units on the upper and lower edges of each plot (in units of $\tau_{\rm{Prec}}$ on the upper edge, and $\tau_{\rm{Kep}}$ on the lower edge). The long-term modulation evident in both GR simulations occurs on the precession timescale $\tau_{\rm{Prec}}$.
    }
    \label{fig:mdot_comps}
\end{figure}

\begin{figure}
    \centering
    \includegraphics[width=\columnwidth]
    {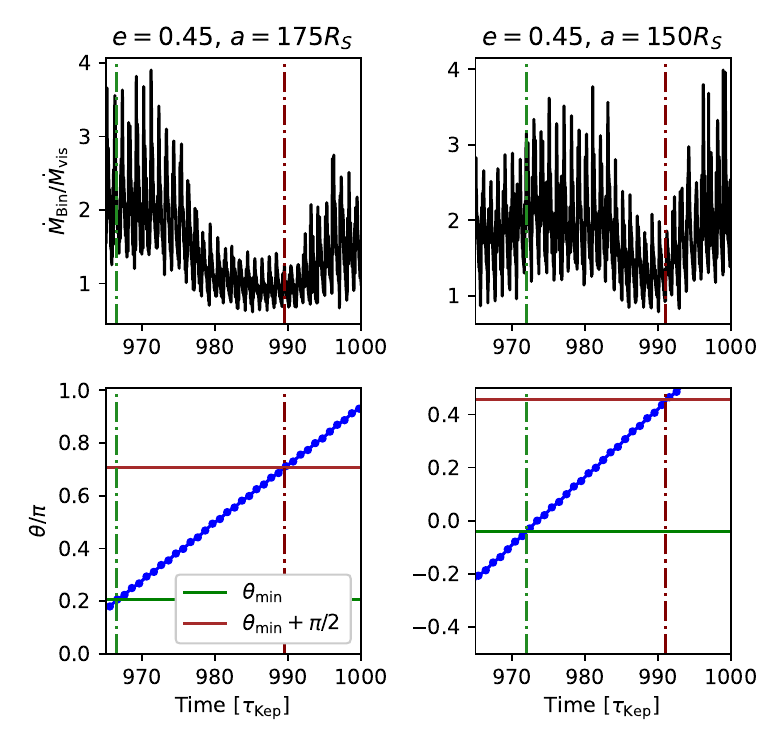}
    \caption{Binary accretion peaks and troughs as the orientation of the binary semi-major axis ($\theta$) rotates with respect to the direction toward the nearest part of the cavity wall ($\theta_{\rm{min}}$). The left and right panels display the $a=175~R_S$ and $a=150~R_S$ simulations, respectively. The upper panels show the total binary accretion rate, while the lower panels plot the angle of the binary's apocenter ($\theta$; blue dots), $\theta_{\rm{min}}$ (green horizontal line), $\theta_{\rm{min}} + \pi/2$ (brown horizontal line); the times at which $\theta$ takes on these two values are indicated using similarly colored vertical dot-dashed lines. The figure shows that in addition to fluctuating on the binary's orbital timescale, the total accretion rate is also modulated on the precession timescale ($0 \leq \theta \leq \pi$).
    The peaks and troughs of this long-timescale modulation occur when the binary's orbit is pointing towards the nearest cavity wall ($\theta=\theta_{\rm min}$) and when it is perpendicular to it ($\theta=\theta_{\rm{min}}+\pi/2$), respectively.} 
    \label{fig:mdot_alignment}
\end{figure}

As illustrated in \autoref{tab:cavity_params}, the cavity center is offset from the binary barycenter in all of our simulations, meaning that the cavities are asymmetric about the BHs. This asymmetry introduces preferential accretion---where material prefers to accrete onto one BH over the other---in our system despite the black holes being the same mass. For a non-circular binary in an asymmetric cavity---in a binary orbit-averaged sense---one of the two BHs will always \footnote{Over long time-scales $>200~\tau_{\rm{Kep}}$ we see a flip in which BH lies closer to the cavity and accretes at a higher rate. This is present in our simulations and has been reported in both \citealt{munoz_lai_16} and \citealt{siwek_pref_accretion}.}
lie closer to the cavity edge than the other, thereby having a stronger tidal pull and higher accretion rate than the other (see \autoref{fig:diagram} and \citealt{siwek_pref_accretion}). This ``preferred" accretion onto one BH over the other is shown in \autoref{fig:mdot_comps}.

\autoref{fig:mdot_comps} shows that our GR and Keplerian simulations display preferential accretion. However, unlike the Keplerian case, the GR simulation does not experience stable preferential accretion. Instead, the accretion rate for each BH varies periodically, with a stable amplitude and a period that is equal to twice the precession period (recall that, in our definitions, the precession period corresponds to the binary precessing by only $\pi$ radians). Further, we note that the two BHs' accretion rates are out of phase with respect to one another. This behavior is due to the binary's precession.

Unlike the Keplerian binary, the GR binary has a precessing apocenter that sweeps out a circle over its precession period (see \autoref{fig:diagram}). Since each apocenter on the circle is a different distance from the cavity wall, the BH's tidal pull on the cavity will vary. As a result, each BH's accretion rate varies quasi-sinusoidally with an angular period of $2\pi$ and thereby a temporal period of $2~\tau_{\rm{Prec}}$. (We remind again that we defined $\tau_{\rm{Prec}}$ as the time it takes to precess by $\pi$, i.e., for the equal-mass BHs to ``switch places".) Further, since the BHs lie directly opposite to one other on the circle, their quasi-sinusoidal fluctuations have a phase difference of $\pi$. The total binary accretion is the sum of the two fluctuating sinusoids with a period of $\pi$, or $\tau_{\rm{Prec}}$, with an amplitude about half that of its components.

To determine when the binary experiences its peak accretion rate, we need to calculate when its tidal interaction with the cavity wall is strongest. We calculate the direction toward the cavity wall segment closest to the binary's barycenter. This direction and distance is represented by the vector $\vec{r}_{\rm{min}}$.

As noted in \autoref{tab:cavity_params}, the barycenters occupy different positions with respect to the cavity's semi-major and semi-minor axes. For example, the barycenter of the $a=175~R_S$ binary is further offset from the cavity-center along the semi-minor axis, and less offset on the semi-major axis than that of the $a=150~R_S$ binary. These different barycenters change $\vec{r}_{\rm{min}}$. Using the values from \autoref{tab:cavity_params} we can calculate $\vec{r}_{\rm{min}}$ explicitly.

In \autoref{fig:mdot_alignment}, we display the binary accretion rate (black), the azimuthal position of a BH's apocenter (blue), and we indicate both the azimuthal direction of $\vec{r}_{\rm{min}}$, denoted as $\theta_{\rm{min}}$, and $\theta_{\rm{min}}+\pi/2$. We note that when the angle of a BH's apocenter matches $\theta_{\rm{min}}$, that BH is closest to the cavity wall and thereby experiences peak accretion. The peak accretion rate of one BH is coincident with peak binary accretion due to the height of the peaks, as is easily inferred from \autoref{fig:mdot_comps}. Similarly, when the binary semi-major axis is perpendicular to $\vec{r}_{\rm{min}}$, binary accretion is at its minimum.

\subsection{Light-curves}\label{ssec:light-curves}

Using the post-processing technique discussed in \autoref{ssec:post_processing}, we generate expected light-curves for our binary systems, calculating the system's bolometric, X-ray, UV, and optical luminosity. In the following section, we describe the amplitude and timing of their peaks, their characteristics in different bands, and their dependence on cavity geometry.

\subsubsection{Long time-scale modulation}\label{sssec:large_scale_signals}
\begin{figure*}
    \centering
    \includegraphics[width=1\textwidth]{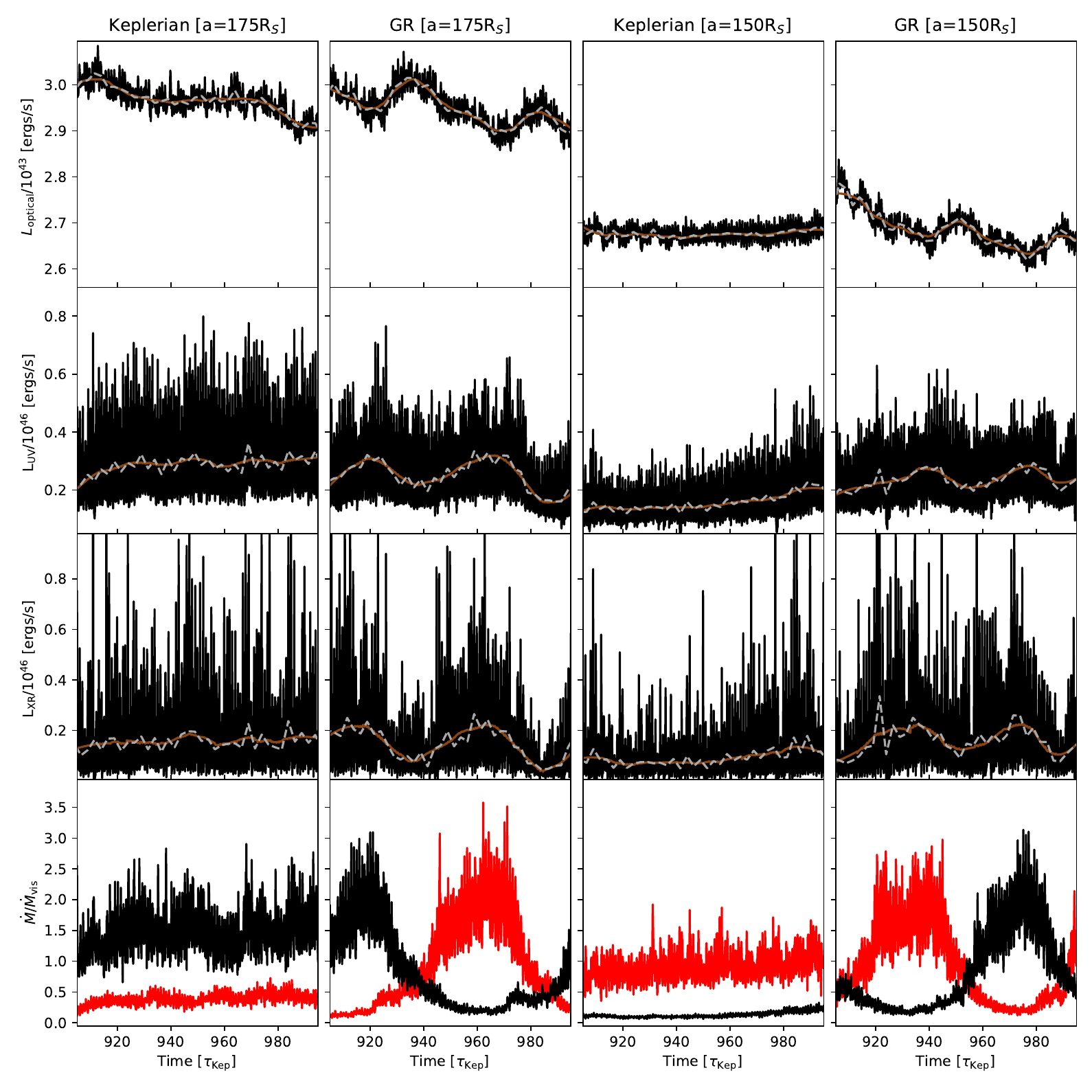}
    \caption{
    Accretion rates and light-curves in our $e=0.45$ simulations. Each column corresponds to the simulation indicated in the column title. In the first three rows, from top to bottom, we plot the optical, UV, and X-ray luminosities. We display the $a=175~R_S$ ($a=150~R_S$) case in the left-most (right-most) two columns. We overlay our calculations of a continuous moving median with a window of $2.5~\tau_{\rm{Kep}}$ (gray) and a Savitsky-Golay filter with a window of $25~\tau_{\rm{Kep}}$ (brown). In the bottom panels, we plot the individual BH accretion rates. Note how both GR simulations show quasi-sinusoidal features that match those of the accretion rates in their X-ray and UV light-curves. The cadence of the data in this plot is standardized to $0.1~\tau_{\rm{Kep}}$.}
    \label{fig:ecc_0.45_masterlightcurve}
\end{figure*}

\begin{figure}
    \centering
    \includegraphics[width=\columnwidth]{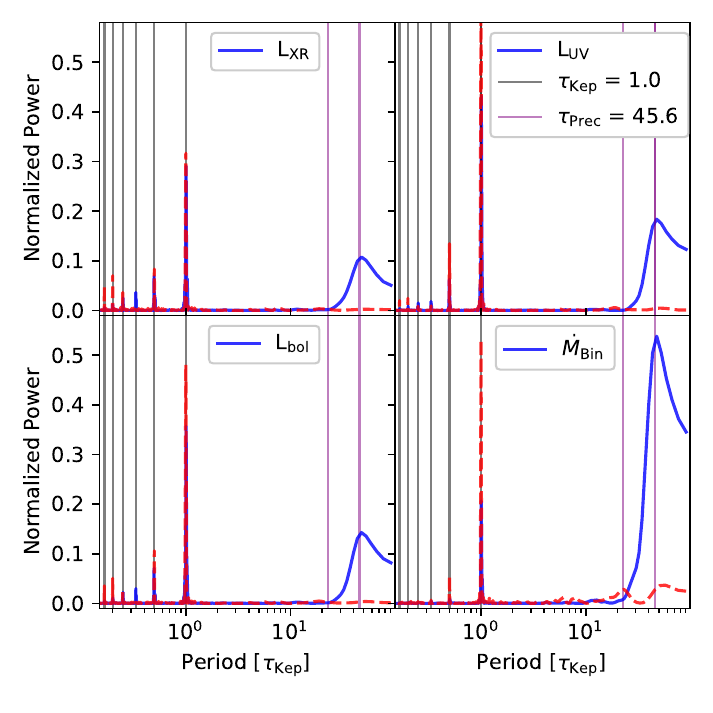}
    \caption{Periodograms for the $e=0.45$, $a=175~R_S$ simulation. The GR and Keplerian versions are represented by the solid blue and dashed red lines, respectively. Clockwise from the lower left, we display the bolometric luminosity, X-ray luminosity, UV luminosity, and binary accretion rate. We also show the harmonics of the different periods: the orbital period of the Keplerian binary (black) and the period of precession $\tau_{\rm{Prec}}$ (purple).
    }
    \label{fig:periodograms}
\end{figure}

In \autoref{fig:ecc_0.45_masterlightcurve}, we show the accretion rates and light-curves for our $e=0.45$ simulations. First, we note that the X-ray and UV light-curves clearly track the behavior of the binary's total accretion rate. The GR simulations exhibit simultaneous peaks and troughs, while the Keplerian simulations display simultaneous small upticks in the moving averages of $\dot{M}_{\rm{bin}}$, $L_{\rm{UV}}$, and $L_{\rm{XR}}$. 

Though both simulations show that the X-ray and UV light-curves track binary accretion, we note that the GR and Keplerian simulations have notably different accretion rates and, thereby, different light-curves. We further delineate the characteristics of these signals by studying their Lomb-Scargle periodograms for the $e=0.45$, $a=175~R_S$ case in \autoref{fig:periodograms}.

In \autoref{fig:periodograms} we see two main peaks across all panels. The first is at $1~\tau_{\rm{Kep}}$ and $1.005~\tau_{\rm{Kep}}$ for the Keplerian and GR simulations, respectively. This peak is the orbital period of each binary. Though the orbital period for the GR binary ($\tau_{\rm{GR}}$) has a slightly longer period than its Keplerian counterpart---due to apsidal precesion---the periods are nearly indistinguishable in \autoref{fig:periodograms} and we only plot the Keplerian orbital period $1~\tau_{\rm{Kep}}$. The orbital period and its harmonics were previously reported in light-curves from non-precessing simulations (e.g.~\citealt{ryan_sailfish}). However, unlike previous simulations, we note that \autoref{fig:periodograms} displays another peak at the precession period. This is a distinct signal unique to the GR case, at $\approx 45.6 ~\tau_{\rm{kep}}$. We note that it is only second in power to the orbital frequency, making the possibility of observing it encouraging (see further discussion in \autoref{sec:observations} below).

While $\dot{M}_{\rm{bin}} (\equiv \dot{M}_1 + \dot{M}_2)$, $L_{\rm{UV}}$, and $L_{\rm{XR}}$ are all similar to each other, $L_{\rm{optical}}$ has distinct behavior. In \autoref{fig:ecc_0.45_masterlightcurve} the optical light-curves for the GR simulation show small amplitude peaks and troughs that are anti-correlated with those of $\dot{M}_{\rm{bin}}$, while those of the Keplerian simulations remain relatively flat. The explanation for the optical light-curves not being correlated with $\dot{M}_{\rm{bin}}$ lies in the high temperature of the mini-disks.

In particular, the mini-disks are the hottest regions in our simulations (see $T_{\rm{eff}}$ map in \autoref{fig:sample_snap}), and thus have black-body spectra that are significantly biased towards the higher energy (X-ray and UV) bands. Thus, only the far tails of their emission are in the optical band. While the heating of the mini-disk causes significant modulations in the X-ray and UV luminosity, these modulations are weakly represented in the optical band. Thus, unlike the higher energy bands, whose light-curve modulations are tied to the gas fed to mini-disks through their interaction with the cavity wall, it is apparent that the flares in the optical light-curves do not. In the following section, we study these flares and the physical mechanism behind the light-curve shapes. 

\subsubsection{Flaring mechanisms}\label{sssec:flaring}

\begin{figure}
    \centering
    \includegraphics[width=1\columnwidth]{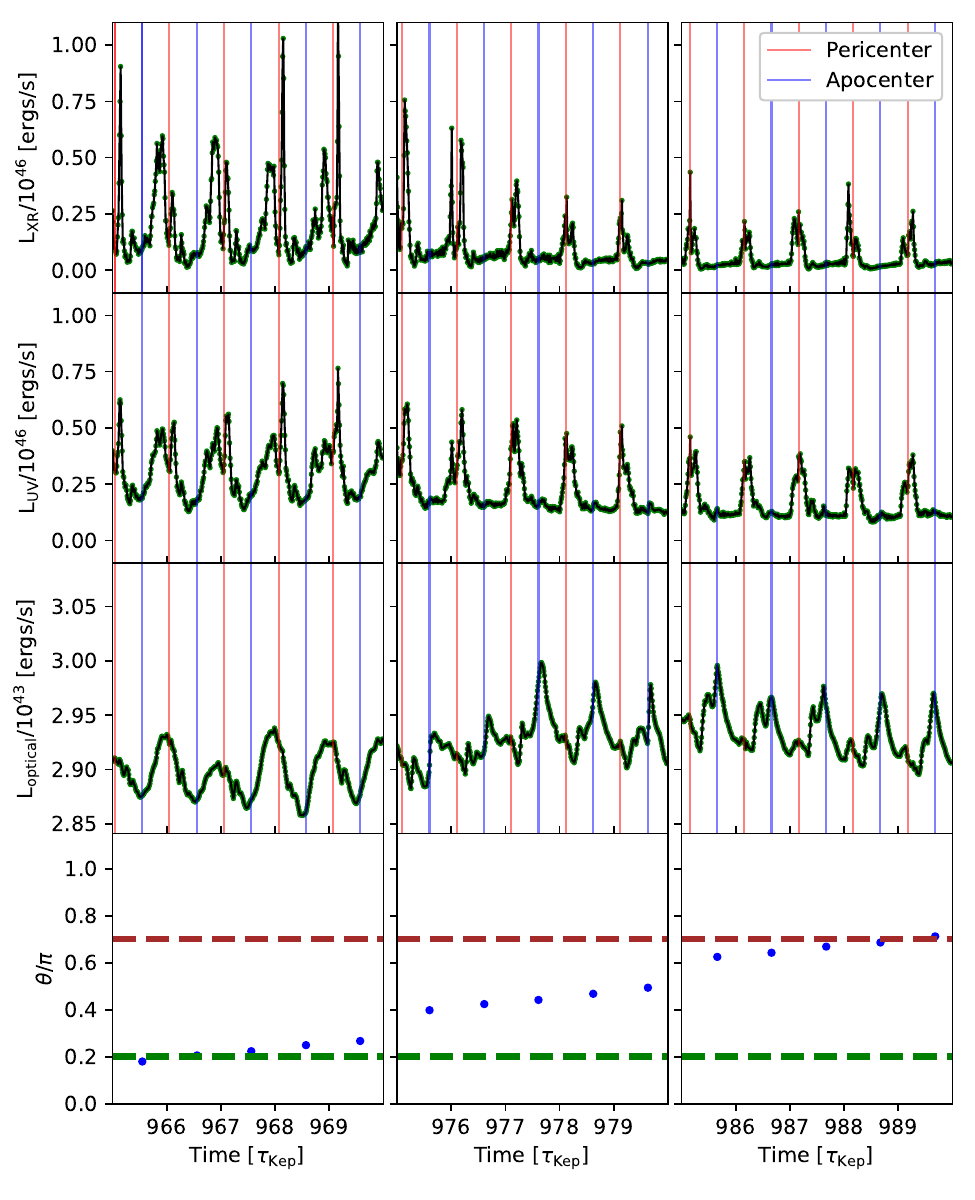}
    \caption{Zoomed-in windows of the $e=0.45$, $a=175~R_S$, GR light-curves. From top to bottom, we display the X-ray, UV, and optical light-curves. The horizontal green and brown dashed lines in the lower panel represent $\theta_{\rm{min}}$ and $\theta_{\rm{min}} + \pi/2$. The blue dots represent the azimuthal position of one BH at apocenter ($\theta$). The cadence of the data is 0.01$~\tau_{\rm{Kep}}$}
    \label{fig:windows}
\end{figure}

\begin{figure}
    \centering
    \includegraphics[width=1\columnwidth]{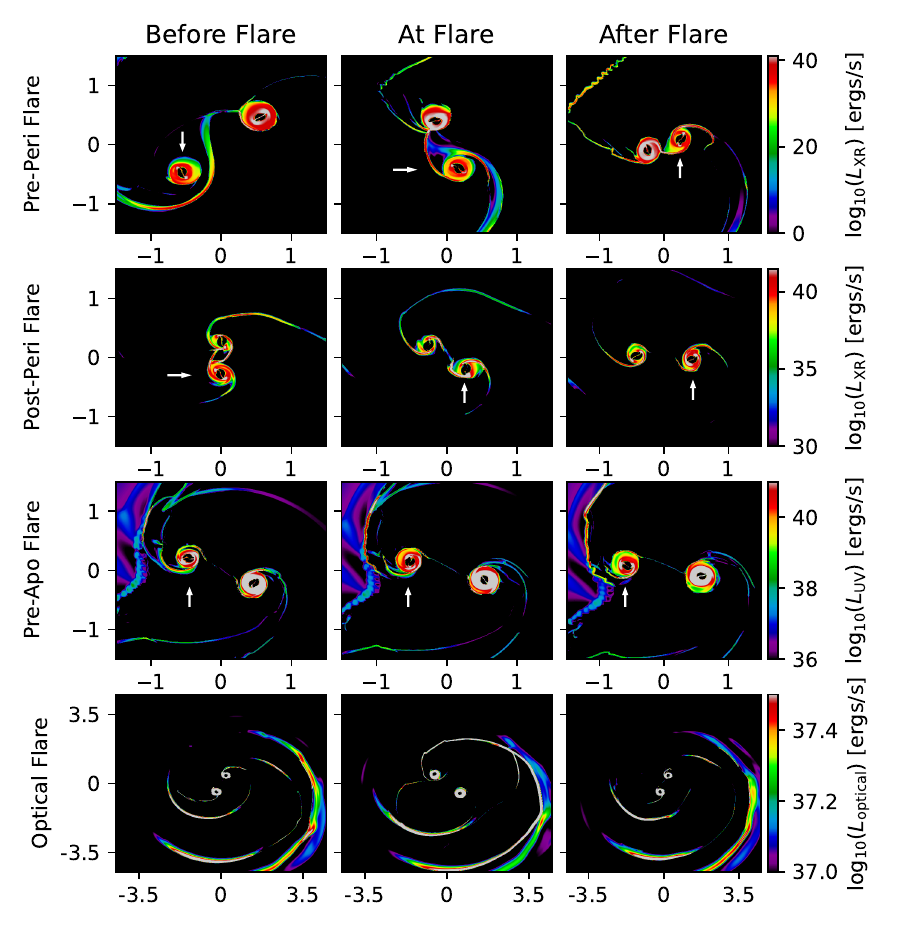}
    \caption{Snapshots clarifying the origin of the four distinct types of flares in our system. From left to right, the columns show luminosity density snapshots just before, at, and just after each flare. The BHs orbit counter-clockwise over time, and the white arrow tracks one of them. The first row of panels shows the X-ray luminosity of our ``pre-peri" flares. The second row displays the X-ray luminosity of our ``post-peri" flares. The third row displays the UV luminosity of our ``pre-apo" flares. The final row displays the optical luminosity of our optical flare.
    }
    
    \label{fig:bandmap_flanks}
\end{figure}

\begin{figure}
    \centering
    \includegraphics[width=1\columnwidth]{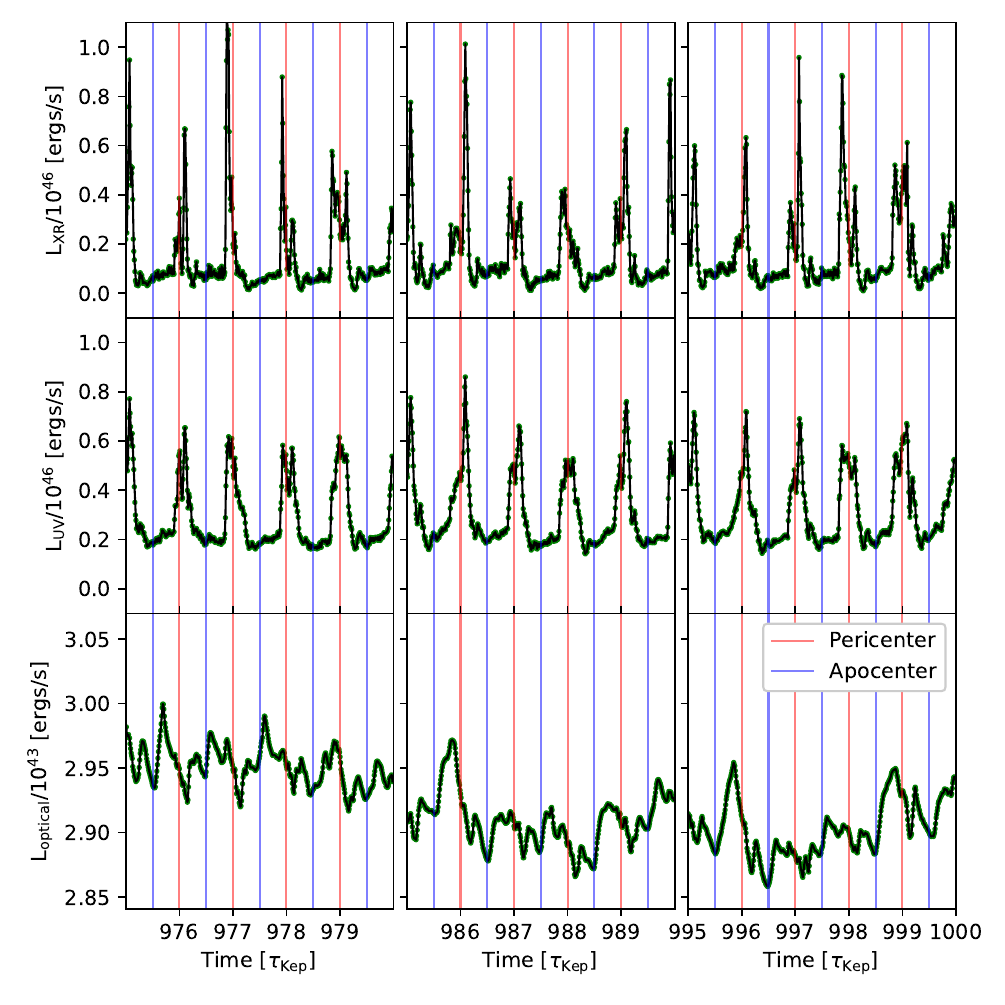}
    \caption{Zoom-in light-curve of the $e=0.45$, $a=175~R_S$ simulation in the Keplerian case. The panels are the same as in \autoref{fig:windows}.}
    \label{fig:windows_kep}
\end{figure}

\autoref{fig:windows} displays zoom-in versions of the X-ray, UV, and optical light-curves for the $e=0.45$, $a=175~R_S$ GR simulation. This figure shows three distinct flares per orbit in the X-ray and UV bands: one occurs just before each pericenter, another just after, and a third one roughly halfway between pericenter and apocenter. We will hereafter refer to these as ``pre-peri", ``post-peri" and ``pre-apo" flares. The optical band, on the other hand, displays a single distinct flare per orbit.

To investigate the physical mechanism behind this particular set of chromatic flares, we examine the system's behavior before and after each of these flares in detail. \autoref{fig:bandmap_flanks} displays a series of snapshots of the surface luminosity density of the system just before, at, and just after each type of flare. The specific flaring instances displayed in \autoref{fig:bandmap_flanks} were arbitrarily chosen and are representative of large-amplitude instances of their respective type of flare.

The first (top) row shows snapshots around the $t=965.92 ~\tau_{\rm{Kep}}$ pre-peri peak. Here, we see two BHs with distinct mindisks and a stream between them. The stream is generated from one BH's interaction with the cavity wall (the BH indicated with an arrow). As the binary begins to approach its pericenter, the counterpart (upper) BH moves closer to the head of the stream, thereby exerting an increasingly strong gravitational pull on the stream. The head of the stream makes contact with the upper BH's mini-disk, depositing enough gas to heat it significantly above the baseline temperature and luminosity, producing a flare. The top row middle panel of \autoref{fig:bandmap_flanks} highlights this deposition, illustrating how the luminous, heated material of the stream is being funneled onto the counterpart (upper) BH.

The second row shows snapshots around the $t=976.18 ~\tau_{\rm{Kep}}$ post-peri peak. These reveal a tidal stripping of the mini-disks, somewhat analogous to Roche lobe overflow. Namely, we see the development of two streams reaching from one mini-disk to the other, forming a `double bridge'. As the BHs orbit, the two streams come closer together until they collide and form a single tidal bridge between the two BHs. This collision effectively squeezes the bridged material, pushing it towards each mini-disk, and causing a flare due to shock heating.

The third row shows snapshots around the $t=966.27 ~\tau_{\rm{Kep}}$ pre-apo flare. These show the BH closer to the cavity wall (left) colliding with the low-luminosity wall. The resulting shock heats the BH's tidal tail and the cavity wall, increasing the system's luminosity and producing the flare.

The fourth and final row shows the snapshots around the time of our optical flare, notably the one at $t=991.70~  \tau_{\rm{Kep}}$. As before, we note that the mini-disks' luminosity do not strongly evolve throughout the flare. Instead, we observe a stream-cavity collision that causes the inner edge of the cavity wall to momentarily shine rather brightly, producing the optical flare we observe.

These mechanisms repeat and produce similar flares in each orbit. However, the amplitude and timing of these flares are not constant from orbit to orbit. 

For example, the pair of pre- and post-peri flares decrease in amplitude and occur closer to each other in time as the binary's accretion rate falls. As discussed above, this happens as the binary's semi-major axis becomes more perpendicular to $\vec{r}_{\rm{min}}$. This is evident when we compare flares from the first and third columns of \autoref{fig:windows}. In the first column, the pre-peri and post-peri flares occur at $t=966.89~\tau_{\rm{Kep}}$ and $967.10~\tau_{\rm{Kep}}$, a separation of $\Delta t =0.21~\tau_{\rm{Kep}}$. The third column, however, shows the pre-peri and post-peri flares at $t=985.16~\tau_{\rm{Kep}}$ and $985.24~\tau_{\rm{Kep}}$, a much smaller separation of $\Delta t =0.08~\tau_{\rm{Kep}}$. Furthermore, the flares during high binary accretion are significantly brighter than those at low binary accretion. We note that the difference between the maximum luminosity of the two aforementioned pre-peri flares is $\approx \Delta \rm{L_{\rm{XR}}} = 2\times10^{45} ~ergs/s$. These effects are due to the changing tidal interaction between the BHs and the cavity, as discussed in the previous section and illustrated in \autoref{fig:diagram}.

More specifically, when the binary's semi-major axis is aligned with $\vec{r}_{\rm{min}}$, one BH is closer to and has a stronger tidal pull on the cavity, funneling significantly more gas to its mini-disk---this creates a large mini-disk and a small one. When the binary is more perpendicular to $\vec{r}_{\rm{min}}$, neither BH is at its closest point to the cavity, and both have more equal tidal interactions with the cavity, leading to mini-disks that are, on average, both small and more equal in size. Furthermore, since the stream---relevant to the pre-peri flare---between the BHs is generated from the interaction of a BH with the cavity wall, it similarly is thinner, harboring less gas, when the binary is more perpendicular to $\vec{r}_{\rm{min}}$.

We find that these differences in the mini-disk and stream size correspond to differences in the timing and amplitude of both the pre-peri and post-peri flares. With a thinner gas stream between the mini-disks, less disrupted stream material needs to travel further before depositing onto the edge of the mini-disk. Furthermore, smaller mini-disks sit deeper inside their host BH’s potential, requiring stronger tidal forces and smaller BH separation to facilitate the formation of a thinner mini-disk bridge. As a result, the amplitude of and time separation between the pair of pre-peri and post-peri flares are reduced as the binary becomes perpendicular to $\vec{r}_{\rm{min}}$. 

We also find long-term variations in the properties of the other two types of flares. The pre-apo flare steadily diminishes in amplitude until it is no longer discernible from the baseline emission - an example of this is seen at $t\approx 989~\tau_{\rm{Kep}}$. The optical flare, on the other hand, both grows in amplitude and shifts its timing from pericenter to apocenter as the binary's orbit becomes more perpendicular to $\vec{r}_{\rm{min}}$. The cause for both of these behaviours can be attributed to the geometry of the system. 

As the binary becomes more perpendicular to $\vec{r}_{\rm{min}}$ the apocenters of both BHs are no longer close enough to the cavity wall to collide with it and produce the pre-apo flare. As for the optical flare, we first note that tidal streams are launched perpendicular to the binary's semi-major axis. Thus, when the binary is perpendicular to $\vec{r}_{\rm{min}}$, the streams propagate in the direction parallel to $\vec{r}_{\rm{min}}$. As a result, the streams travel a shorter distance and are less diffuse when they collide with the cavity, causing brighter flares closer to the previous pericenter from which they were launched.
 
Finally, we turn to \autoref{fig:windows_kep}, where we show the detailed light-curve of the Keplerian $e=0.45$, $a=175~R_S$ simulation. As the figure reveals, we see the same types of flares in the X-ray and UV. They occur at the same points during the binary's orbit and are caused by the same physical mechanisms. However, there is no significant change in the amplitude or timing of these flares in comparison to the GR case displayed in \autoref{fig:windows}.

Indeed, the phenomena of tidal stripping, the formation of tidal streams, stream-cavity, and stream-mini-disk collisions are not unique to a precessing binary. Instead, they are generic to an eccentric binary that strongly interacts with an asymmetric cavity. On the other hand, the modulation in the amplitudes and timings of these flares are uniquely caused by the modulations in the accretion rate coming from the precession of the binary's semi-major axis. Thus, the modulation in the amplitudes and timings of these flares is a GR effect.

\section{Near-circular binaries}\label{sec:weak_gr}

In the previous sections, we discussed the effects of GR precession in our $e=0.45$ simulations, detailing its signatures on accretion rates and light-curves. One expects such signatures to change qualitatively for sufficiently smaller eccentricity. In this section, we show that this qualitative change has already occurred by $e=0.15$.

In the top panels of \autoref{fig:ecc_0.15_accretion} we display the accretion rates in our $e=0.15$ simulations. Many of the characteristic signatures detailed in \autoref{ssec:accretion_rate} are missing. Rather, in both the GR and Keplerian runs, we see comparatively mild preferential accretion and a binary accretion rate that has a starkly different time dependence in comparison with the $e=0.45$ binaries considered in previous sections. This accretion behavior and the corresponding features of the resulting light-curves are determined by the binary cavity geometry and the non-trivial evolution of the cavity itself. In the following sections, we detail these unique features in our $e=0.15$ simulations.

\subsection{Binary-cavity interaction}\label{ssec:damped_pref}
    
\begin{figure}
    \centering
    \includegraphics[width=1\columnwidth]{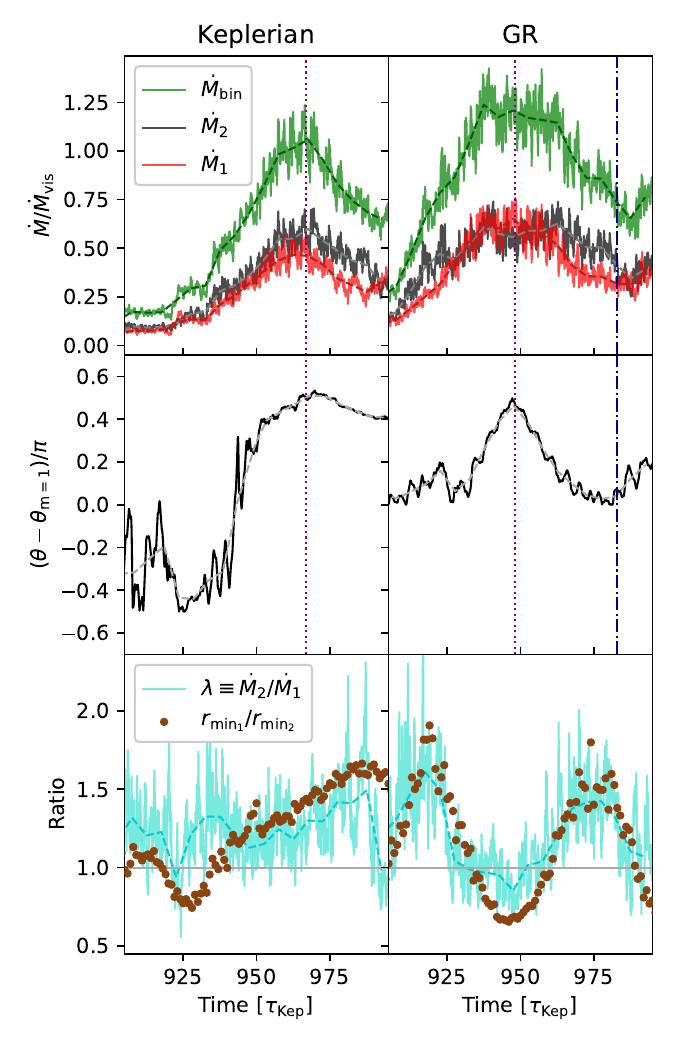}
    \caption{We display data from our $e=0.15$ Keplerian (left) and GR (right) simulations. The top row shows the individual ($\dot{M}_1$, $\dot{M}_2$) and total ($\dot{M}_{\rm{bin}}$) accretion rates in red, black, and green, respectively. In the middle row, we show the azimuthal angle between the binary's semi-major axis and the $m=1$ mode of the inner circumbinary disk (see \autoref{ssec:cavity_precession}), $\theta_{\rm{rel}} \equiv \theta - \theta_{\rm{m=1}}$ in black. The bottom panels show the accretion ratio $\lambda \equiv \dot{M}_2/\dot{M}_1$ in turquoise and the ratio of the minimum distances from each BH to the cavity (as fit by an ellipse; see \autoref{sec:morphology}) $r_{\rm{min}_1} / r_{\rm{min}_2}$ in brown. The cadence of $r_{\rm{min}_1} / r_{\rm{min}_2}$ is $1~\tau_{\rm{Kep}}$, while all other data has a cadence of $0.1~\tau_{\rm{Kep}}$. We overlay moving medians with a window of $5 ~\tau_{\rm{Kep}}$ of each dataset with similarly colored dashed lines. We plot purple (dotted) and blue (dash-dotted) vertical lines where $\theta_{\rm{rel}}$ is a local maximum and minimum, respectively.}
    \label{fig:ecc_0.15_accretion}
\end{figure}

The first row in \autoref{fig:ecc_0.15_accretion} displays the accretion rates in our two $e=0.15$ simulations, depicting $\dot{M}_{\rm{bin}}$, $\dot{M}_2$, $\dot{M}_1$ in green, black, and red respectively. We see that the accretion rates on both BHs are fluctuating much less than for their $e=0.45$ counterparts. In fact, the $e=0.45$ BHs can accrete up to $3$ times the viscous inflow rate $\dot{M}_{\rm{vis}}$, which is $\approx 5$ times the maximum for the $e=0.15$ BHs---which only accrete up to $\approx 0.6 ~\dot{M}_{\rm{vis}}$ each. The difference between the accretion rates of the two BHs in the $e=0.15$ case is also smaller than in the $e=0.45$ case. Where $|\dot{M}_1-\dot{M}_2|$ was $\gtrsim 1 \dot{M}_{\rm{vis}}$ for $e=0.45$ binaries, it is $\approx 0.1 \dot{M}_{\rm{vis}}$ for $e=0.15$ binaries: an order of magnitude decrease. This reduction in accretion fluctuation and the accretion differential $|\dot{M}_1-\dot{M}_2|$ is chiefly due to the change in the binary's apocenter with respect to the cavity.

As per \autoref{tab:cavity_params}, we note that the ellipses fit to the $e=0.15$ simulations' cavities have centers that are much closer to the binary barycenter than that of the $e=0.45$ simulations, meaning that they are more symmetric about the barycenter. This reduces the differential between each BH's distance to the cavity and, thereby, the differential in accretion rates. Further, binary apocenters scale with $e$: the $e=0.45$ and $e=0.15$ binaries have apocenter values of $1.45~a$ and $1.15~a$, respectively. Since a reduction in the radial extent of the binary increases the distance from both BHs to the cavity, there is a decrease in the maximum accretion rate--the amplitude of the fluctuation.

While the more circular shape of the binary orbit and the more central position of its
apocenter help explain the $e=0.15$ simulations' smaller accretion fluctuations and accretion differentials, they do not illuminate their particular time evolution. To do so, we must account for the cavity's precession.

\subsection{Cavity precession}\label{ssec:cavity_precession}

An implicit assumption in \autoref{sec:strong_gr} is that the orientation and shape of the cavity and the CBD are constant in time. However, the CBD can have a precession of its own. Recent studies have elucidated how CBD precession depends on binary eccentricity \citep{miranda_munoz_lai_17,siwek_pref_accretion, tiede_disk_prec}.

To characterize the precession of the cavity, we calculate the complex moment of the surface density
\begin{equation}\label{eqn:fourier_transform}
     \int_{0}^{2\pi} \Sigma(r, \theta) e^{im\theta} \,d\theta
\end{equation}
where we pick out the $m=1$ mode and define $\theta$ to have an origin about the barycenter. We compute the integral in \autoref{eqn:fourier_transform} over a radial annulus with an inner edge of $r=1~a$ and an outer edge of $r=6~a$ to fully capture the behavior of the asymmetric cavity and not be biased by behaviors in the outer disk.  We have checked that our results are not sensitive to the precise definition of these boundaries.

The phase of the complex number defined in \autoref{eqn:fourier_transform}, which we denote by $\theta_{m=1}$, is used to characterize the inner CBD's orientation and precession. The CBD's orientation at a given time is given by the predominant value of $\theta_{m=1}$ around that time, whereas the CBD's precession is given by a long-term linear trend in $\theta_{m=1}$.\footnote{This phase $\theta_{m=1}$ is also a sensitive measure of $m=1$ waves propagating wave along the cavity wall. For example, the fast oscillations in the middle row of \autoref{fig:ecc_0.15_accretion} have a period of roughly five orbits, and thus are likely the remnants of a ``lump'' phenomenon that is too weak to imprint noticeably on accretion rates, but can still be detected using the sensitive phase diagnostic $\theta_{m=1}$ \citep[for examples of weak lump-like phenomena for eccentric binaries embedded in CBDs, see][]{ryan_sailfish}.} We do this for each simulation snapshot, developing a time series of the mode's azimuthal position. 

The above procedure was performed on all our simulations. We found that while all our cavities experienced some precession, that of our $e=0.45$ simulations was insignificant. The highest cavity precession rate in our $e=0.45$ simulations was only $0.006 \  \rm{radians}/ \tau_{\rm{Kep}}$ and thus did not strongly affect our results. However, both $e=0.15$ simulations experience much faster cavity precession rates, with the highest cavity precession rate in our $e=0.15$ simulations being $\approx 0.11 \ \rm{radians}/\tau_{\rm{Kep}}$---over an order of magnitude increase. Despite having different thermodynamics, these results agree with previous models of Keplerian binary accretion by \citealt{miranda_munoz_lai_17} and \citealt{tiede_disk_prec}, which report that $e=0.45$ binaries have apsidally locked disks (i.e.~they do not precess), and that $e=0.15$ binaries display strong disk precession. While a systematic study on how binary parameters affect the cavity's precession remains beyond this paper's scope, we will detail next how cavity precession affects accretion and the associated light-curves.

\subsection{Binary accretion}\label{ssec:circ_binary_accretion}

In \autoref{ssec:accretion_rate}, we illustrated how the alignment between $\vec{r}_{\rm{min}}$ and the binary's semi-major axis affects the accretion rate. We found that in a practically non-precessing cavity, when the binary semi-major axis aligns (anti-aligns) with $\vec{r}_{\rm{min}}$, the binary accretion rate is at a maximum (minimum). When we account for the precession of the cavity, we find that the total accretion rate of the $e=0.15$ binary also depends on the binary's relative position to the cavity.

We note that at each time step, the complex $m=1$ mode of the surface density defines a vector that points to the segment of the disk that is closest to the barycenter, essentially serving as a proxy for $\vec{r}_{\rm{min}}$. Thus, in the middle panels of \autoref{fig:ecc_0.15_accretion} we present the azimuthal position of the binary's semi-major axis with respect to that of $\theta_{m=1}$, an angle we will call the relative binary-cavity angle: $\theta_{\rm{rel}}$.

It is clear from the figure that binary accretion closely tracks $\theta_{\rm{rel}}$, with the two rising and falling together. For both the GR and the Keplerian simulation, we see that when $\theta_{\rm{rel}}$ is at a local maximum or minimum, so are the binary accretion rates. The purple (dotted) and blue (dash-dotted) vertical lines plotted in \autoref{fig:ecc_0.15_accretion} 
highlight the coincidence of the peaks and valleys, respectively.

We further note that the accretion rates peak when $\theta_{\rm{rel}} \approx \pi/2$ and are lowest when $\theta_{\rm{rel}} \approx 0$. Equivalently, accretion is highest (lowest) when the binary semi-major axis and $\vec{r}_{\rm{min}}$ are anti-aligned (aligned). This is the opposite of our trend in \autoref{ssec:accretion_rate}.

The behaviour in \autoref{ssec:accretion_rate} had an intuitive explanation: whenever the binary's semi-major axis aligns with $\vec{r}_{\rm{min}}$, one BH is closer to the cavity and thus the accretion rate of this BH spikes, while that of the other, more distant BH dips. If this were the dominant effect, then precession should cause the {\it out-of-phase} ``see-saw" oscillations between $\dot{M}_1$ vs. $\dot{M}_2$ exhibited by the GR simulations in \autoref{ssec:accretion_rate}. However, for our much less eccentric $e=0.15$ binary, the accretion rates of the two BHs clearly vary {\it in tandem}, rather than out-of-phase.  How can the behaviour be so different?

First, we note that for the above, the purely geometric modulation illustrated in \autoref{fig:diagram} should occur for any value of the binary eccentricity. The amplitude of the see-sawing should scale monotonically with $e$ and become much less prominent in the $e=0.15$ case to the extent that the two accretion rates in the top panels of \autoref{fig:ecc_0.15_accretion} are not strongly differentiated.

Second, we point out that while the $e=0.15$ cavity is more symmetric than its $e=0.45$ counterpart, it is still strongly lopsided, and its precession is not necessarily geometrically equivalent to the binary precession shown in \autoref{fig:diagram}. The precise shape of the cavity may evolve, and even for a fixed cavity shape, the precession may occur about a point that is offset from the barycenter. While we could not unambiguously diagnose such an offset, our results do suggest that as the cavity precesses, the closest distance from the binary's barycenter to the cavity wall oscillates, modulating the accretion rates onto both BHs together. Furthermore, this modulation is apparently far dominant over the significantly dampened out-of-phase oscillations caused by the binary's precession.  

Finally, we note that binary accretion rates are largest when both BHs are roughly equidistant from the cavity wall and accreting close to equally (i.e. at $\theta_{\rm{rel}} \approx \pi/2$), and are smallest when one BH is closer to the cavity wall than the other and accreting unequally (i.e. at $\theta_{\rm{rel}} \approx 0$).

\subsection{Accretion ratio}\label{ssec:circ_accretion_ratio}

In addition to the accretion ratio, we also show the ratio of the minimum distances from each BH to the cavity wall ($r_{\rm{min}_1} / r_{\rm{min}_2}$) in the bottom row of \autoref{fig:ecc_0.15_accretion}. Namely, we calculate $r_{\rm{min}}$ for each BH using the ellipses fit to the $10^{-3} ~M_{\rm{bin}}a^{-2}$ surface-density contour, as originally described in \autoref{sec:morphology}. Despite the fact the cavity is not a perfect ellipse, which inherently limits the usefulness of $r_{\rm{min}}$, we find that the inverse ratio of the distances from the BHs to the cavity wall does, in fact, trace the general features in the evolution of the accretion ratio. As one BH gets closer to the cavity wall than the other, it accretes more than the other.

While low eccentricity binaries tend to have both BHs accrete relatively in phase, their accretion ratio is interestingly still dependent on the BHs' relative distance to the nearest cavity wall. Thus, the geometric mechanism detailed in \autoref{fig:diagram} is still present in the $e=0.15$ binary. However, the effect is significantly damped by smaller apocenter values.

\subsection{Beat frequency}\label{ssec:beat}

\begin{figure}
    \centering
    \includegraphics[width=1\columnwidth]{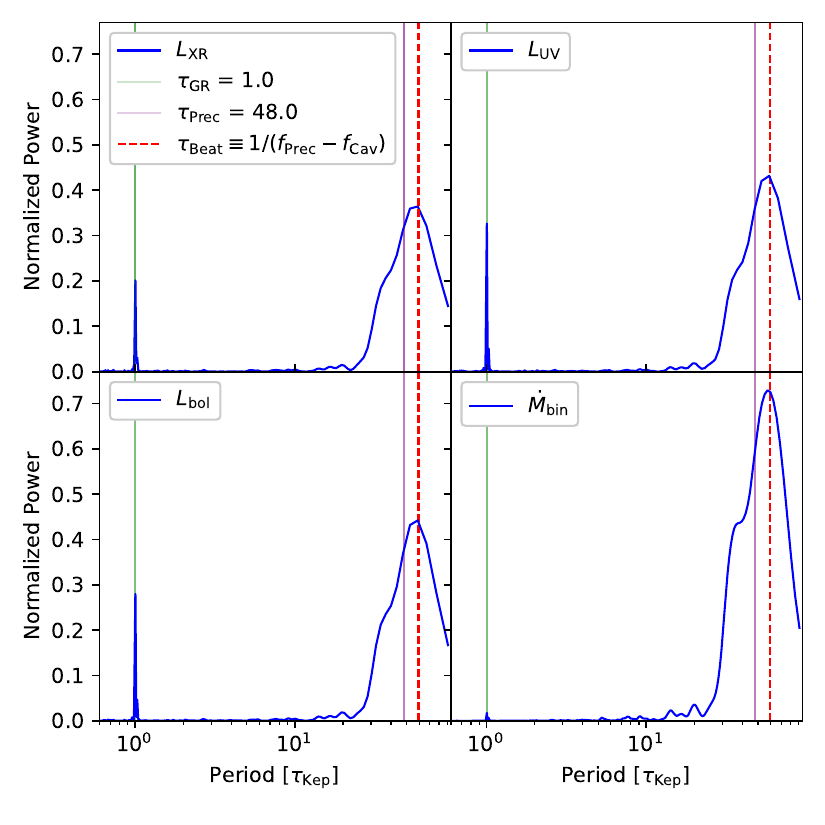}
    \caption{Periodograms in the $e=0.15$ GR simulation. Clockwise from the lower left, the four panels display periodograms of the bolometric luminosity, X-ray luminosity, UV luminosity, and the binary accretion rate. We plot $\tau_{\rm{Prec}} \equiv f_{\rm{{Prec}}}^{-1}$ (purple solid line), the orbital period of the GR binary $\tau_{\rm{GR}}$ (green solid line), and the beat between the period for the cavity to precess by $2\pi$ ($\tau_{\rm{Cav}}=f_{\rm{Cav}}^{-1}$) and the GR binary to precess by $\pi$ ($\tau_{\rm{Prec}}$) $\tau_{\rm{Beat}}=1/(f_{\rm{Prec}}-f_{\rm{Cav}})$ (red dashed line). We find that the beat is the dominant period in the accretion rate and light-curves.}
    \label{fig:periodograms_circle}
\end{figure}

The $e=0.15$ GR simulation exhibits two precession periods: the apsidal precession of the binary and the precession of the cavity. The interplay between these precessions not only affects the accretion of the binary as shown in \autoref{fig:ecc_0.15_accretion}, but having two competing precession effects also imparts a dominant beat frequency into the accretion rate and associated light-curve signal.

In \autoref{fig:periodograms_circle} we show the periodograms of $L_{\rm{XR}}$, $L_{\rm{UV}}$, $L_{\rm{bol}}$, $\dot{M}_{\rm{bin}}$ for our $e=0.15$ GR simulation. As before, we note that our system's bolometric and high-energy bands are tied to the behavior of the mini-disks and thereby have similar periodograms to $\dot{M}_{\rm{bin}}$. However, unlike \autoref{fig:periodograms}, we note that the luminosity periodograms do not peak around the binary precession period, but rather show a peak at the beat ($\tau_{\rm{Beat}}=1/(f_{\rm{Prec}}-f_{\rm{Cav}}$) between the binary precession period ($\tau_{\rm{Prec}} \equiv f_{\rm{Prec}}^{-1}$) and the period for the cavity to precess by $2\pi$ ($\tau_{\rm{Cav}} \equiv f_{\rm{Cav}}^{-1}$). Namely, from  $\theta_{\rm{m=1}}$ we determine that the period for the cavity to precess by $2\pi$ is $\approx 250~\tau_{\rm{Kep}}$, which is quite similar to the values found by \citealt{siwek_pref_accretion} and \citealt{miranda_munoz_lai_17} in the isothermal case. We note that both precessions are prograde, and it is expected that the peak frequency would be the beat frequency since the beat frequency is simply the inverse of the cavity precession period if we were in the co-rotating frame of the binary.
We also remind the reader that we defined the binary precession period $\tau_{\rm{Prec}}$ to correspond to precession by $\pi$ radians, since for the equal-mass binaries we consider, the binary looks the same after half a rotation. For sufficiently unequal-mass binaries, one might instead expect a full $2\pi$ precession of the binary to imprint on the dynamics.

\section{Observability}\label{sec:observations}

In this section, we determine the parameter space where we could realistically hope to observe GR precession of the binary in EM and/or GW data. Namely, we require such systems to satisfy three constraints: (i) there are significant interactions between an eccentric binary and its CBD cavity, (ii) the binary has a sufficiently long GW-driven inspiral that is not exceedingly rare, and (iii) precession is fast enough for detection.

\subsection{Cavity requirements}\label{ssec:cavity_req}

\begin{figure}
    \centering
    \includegraphics[width=1\columnwidth]{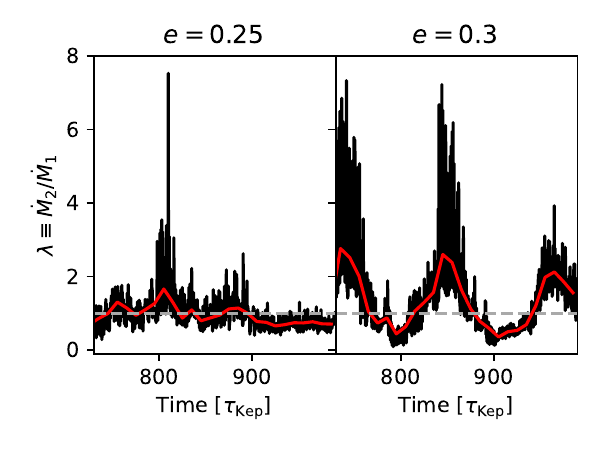}
    \caption{Illustration of the sharp change in accretion behavior when 
    the binary begins to interact strongly with the cavity wall. In the left panel, we plot the accretion ratio $\lambda \equiv \dot{M}_2/\dot{M}_1$ for the $e=0.25$ (left) and the $e=0.3$  (right) case in black (with semi-major axis $a=150~R_S$ in both cases). In red, we overlay a moving median with a window of $10~\tau_{\rm{Kep}}$. The cadence of the data is $0.1~\tau_{\rm{Kep}}$ and the simulations are evolved as prescribed in \autoref{ssec:numerics}.}
    \label{fig:sweep_diptych}
\end{figure}

Much of the characteristic precession-based EM signatures we have detailed in \autoref{ssec:light-curves} are dependent on strong interaction between the binary and the cavity wall (i.e large-amplitude quasi-sinusoidal $\lambda$). The extent to which a binary interacts with the cavity wall depends on the choice of $e$ rather than $a$. To determine the $e$ space where the binary interacts with the cavity strongly, we ran $14$ GR simulations with $a=150~R_S$ and $e \in [ 0.05, 0.7 ] $ with a step size of $0.05$. We evolve each simulation as described in \autoref{ssec:numerics}.

The accretion ratio $\lambda$ is a sensitive diagnostic that can reveal subtle differences in the BH accretion rates. For example, $\lambda$ indicates a sharp change in accretion behavior from $e=0.25$ to $e=0.3$. In \autoref{fig:sweep_diptych}, we plot the accretion ratio of both simulations in black, overlaying the moving median with a window of $10 ~\tau_{\rm{Kep}}$ in red. The differences between the panels in \autoref{fig:sweep_diptych} are stark. The $e=0.25$ case does not display any periodicity, nor does the accretion ratio fluctuate significantly away from unity. The $e=0.3$ case, however, shows a clear large-amplitude oscillatory behavior. In fact we note that, as per \autoref{ssec:accretion_rate}, the period of its modulation is close to $2~\tau_{\rm{Prec}}$, where $\tau_{\rm{Prec}} \approx 44.6 ~\tau_{\rm{Kep}}$. 

This pair of examples represents a genuine transition: the accretion ratio $\lambda$ for binaries with $e \leq 0.25$ are similar, and do not display a large-scale period or amplitude. The $\lambda$ for simulations with eccentricities $e \geq 0.3 $, however, all show periodicities on the time-scale of $2 ~\tau_{\rm{Prec}}$  and with amplitudes monotonically increasing with $e$---presumably due to the monotonically increasing value of their apocenters. Thus, we set $e=0.25$ as an eccentricity floor, at or below which binary eccentricity no longer interacts sufficiently strongly (for our purposes) with the cavity wall.

\subsection{Timescale requirements}\label{ssec:timescales}

\begin{figure*}
    \centering
    \includegraphics[width=1\textwidth]{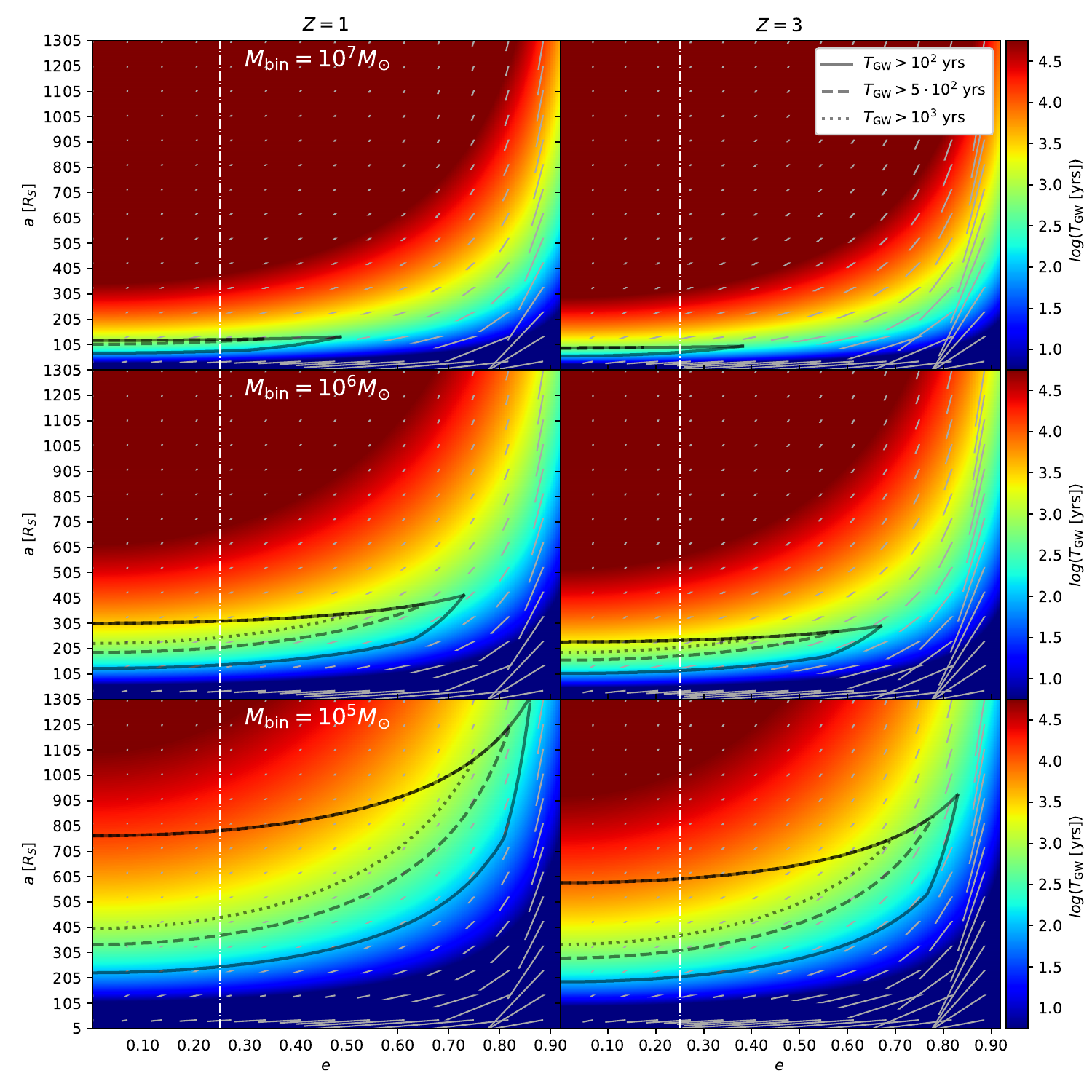}
    \caption{Parameter space where the EM signature of a binary experiencing GR precession could be observable. We represent the parameter space for equal-mass binaries with total masses of $10^{7}~M_{\astrosun}$, $10^{6}~M_{\astrosun}$, and $10^{5}~M_{\astrosun}$ (top to bottom) at $z=1$ (left) and $z=3$ (right). Each panel shows the same region of $a$,$e$ parameter space, color-mapped to the time to merger ($\tau_{\rm{GW}}$) in years. The black curves enclose the region in which the binary system satisfies the timescale constraints of \autoref{ssec:timescales}, with the upper curve marking $\tau_{\rm{Prec}}=3$ years, and the lower curve marking the various $\tau_{\rm{GW}}$ constraints. The vertical white dot-dashed line at $e=0.25$ marks the binary eccentricity threshold for significant preferential accretion, as discussed in \autoref{ssec:cavity_req}. The grey line segments indicate the GW-driven decrease in the binary's $e$ and $a$ over 10 precession cycles.}
    \label{fig:heatmap}
\end{figure*}

At the late stages of MBHB inspiral, the GW-driven inspiral outpaces the viscous timescale in the surrounding CBD. This was suggested, based on semi-analytic models, to lead to a ``decoupling" of the binary from the disk~\citep{Liu+2003-decouple,Milos+2005-decouple,haiman_09}. Around this nominal decoupling time, two distinct effects may be expected: the binary may stop accreting from the disk, and the orbital evolution of the binary becomes dominated by GWs, rather than the gas disk. Recent hydrodynamical simulations~\citep{farris_2015_2,Tang+2017-decouple,dittmann_decouple,luke_decoupling,Franchini+2024-decouple} addressed the first of these effects, and found that most binaries can accrete and remain luminous well past decoupling. Thus we do not consider the scenario where the binary no longer accretes from the disk. On the other hand, there remains an uncertainty about when and how the surrounding disk influences the orbital evolution of the binary. 

As mentioned in \autoref{sec:intro}, recent hydrodynamical simulations have also shown that a circumbinary disk induces an eccentricity of $e \approx 0.4-0.5$  in the binary's orbit \citep{munoz_miranda_lai_19, zrake_eqbm_ecc, duffell_dorazio_2020, siwek_orbevol}. These simulations have not explicitly included the impact of GWs, i.e., they assumed the system is prior to any decoupling. Assuming that past the nominal decoupling, the GWs indeed dominate the orbital evolution, the orbits will rapidly circularize, reducing and ultimately eliminating relativistic precession effects.\footnote{In the presence of misaligned spins, there could still be precession, which we do not address in this paper.}

Thus, to be conservative, we do not assume that the circumbinary disk influences our binary evolution, so we only account for GW decay. In light of this, we define timescale constraints that ensure that our binaries are common and long-lived, and that the EM precession signal we have detailed is observable. Namely, we place constraints on (i) the GW time to merger $\tau_{\rm{GW}}(a,e)$, (ii) the number of precession cycles the binary would exhibit in its lifetime defined by $\tau_{\rm{GW}}(a,e) / \tau_{\rm{Prec}}(a,e)$, and (iii) the length of the precession cycle $\tau_{\rm{Prec}}(a,e)$. 

Specifically, we require $\tau_{\rm{Prec}}(a,e) < 3$ years to observe at least several cycles on a human time scale. We also require $\tau_{\rm{GW}}(a,e) / \tau_{\rm{Prec}}(a,e) >100$ to ensure that the binary would display enough cycles during its lifetime. Finally, we require $\tau_{\rm{GW}}(a,e)$ to be greater than $\{ 10^2, 5\times 10^2, 10^3 \}$ years, to ensure that these binaries are not exceedingly rare (characterised by the ratio of the merger time to quasar lifetime, see \cite{Xin_2021}). The regions of parameter space that satisfy these constraints are enclosed by the black curves in \autoref{fig:heatmap}.

In particular, we plot three curves, with each pertaining to one of the $\tau_{\rm{GW}}(a,e)$ constraint values. The upper portion of these curves reflects the $\tau_{\rm{Prec}}(a,e)$ constraint of $3$ years (binaries need to be below this curve to precess fast enough). The lower portion of these curves reflects the $\tau_{\rm{GW}}(a,e)$ constraint of $\{10^2, 5 \times 10^2, 10^3\}$ years, respectively (binaries need to be above this curve to be common enough). Finally, we find that the condition for $\tau_{\rm{Prec}} \geq 100 ~\tau_{\rm{GW}} $ does not impose a strong constraint, only contributing the steeply upward sloping portion of the lower curve displayed exclusively for high $e$ and low $a$ values. The vertical white dot-dashed lines mark the binary eccentricity threshold of $e=0.25$ for strong interaction with the cavity, as discussed in \autoref{ssec:cavity_req}.

We find that the region of observability expands as the binary mass decreases. This is largely due to requiring the precession timescale to be $\leq 3$ years. When normalized to orbital periods, $\tau_{\rm{prec}}$ only depends on $q$, independent of binary mass at a constant separation $a/R_S$. Since the orbital period, at a constant $a/R_S$, increases with binary mass, we see the region of binaries with $\tau_{\rm{prec}}\leq 3$ years expands dramatically with a decrease in mass. Furthermore, we see that the regions of observability shift slightly towards higher $a/R_S$ as the binary mass decreases. This is because $\tau_{\rm{GW}}(a,e)$ increases with binary mass at a constant separation $a/R_S$. Finally, we note that all regions shift to lower $a/R_S$ with an increase in $z$. 

The conclusion is that for $10^{7}~{\rm M_{\astrosun}}$ binaries at $z=1$, the largest observable region is a very narrow ``sliver", covering $50\lesssim a/R_{\rm S}\lesssim 120$ and $0.25 \leq e\lesssim 0.45$. On the other hand, the chances of observing precession effects for a lower-mass system increase; for $10^{5}~{\rm M_{\astrosun}}$ binaries, the largest observable region covers $200\lesssim a/R_{\rm S}\lesssim 1300$ and $0.2\leq e\lesssim 0.85$. Furthermore, this trend extends outside the three binary masses we have specified in the panels. We find that $\approx 3 \times 10^{7}~M_{\astrosun}$ systems are the most massive binaries that have a non-zero region which satisfies our most generous constraints, while lower-mass systems have continually expanding regions of observability, assuming that such intermediate-mass binaries exist and that they too are embedded in circumbinary disks.

Finally, as noted before, binaries shrink and circularize due to GW-driven decay. The gray line segments in \autoref{fig:heatmap} indicate the change of binary parameters over $10~\tau_{\rm{Prec}}$. While these evolutionary effects in the observable regions are slow enough not to interfere with the characterization of the signal, if the systems were to be observed for longer than 10 precession timescales, we would expect our precession signal's period to lengthen and then, ultimately, disappear.

\subsection{GW detections}\label{ssec:gw_detection}

\begin{figure}
    \centering
    \includegraphics[width=1\columnwidth]{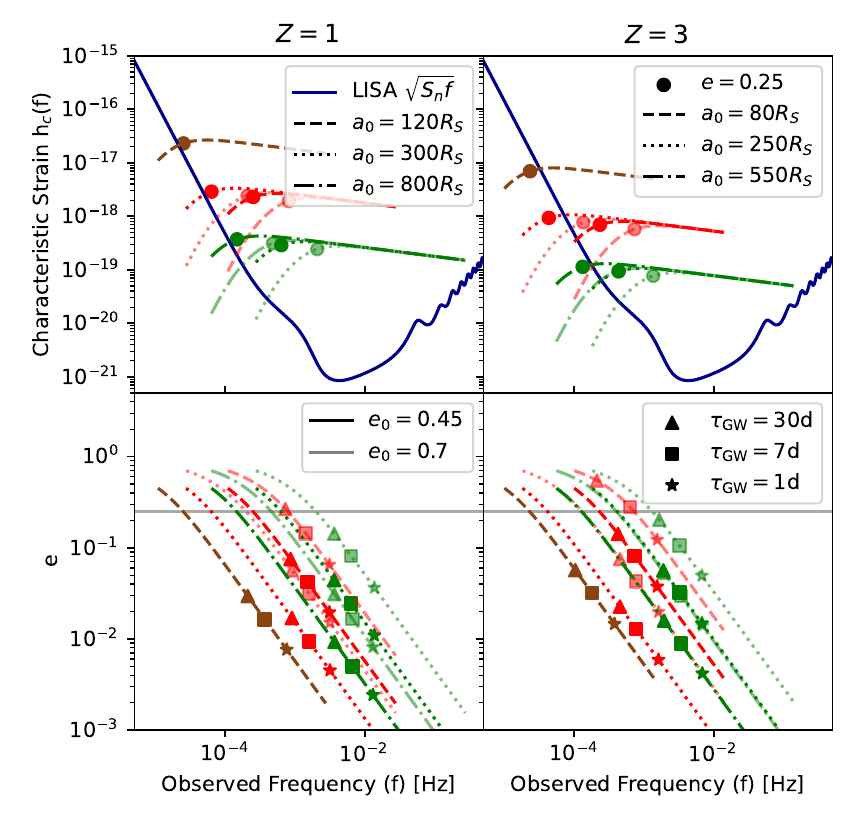}
    \caption{The observability of precessing systems in GWs. The upper and lower panels illustrate the characteristic strain and eccentricity of several different binaries as a function of observed frequency. The left and right panels display results for binaries at $z=1$, and $z=3$, respectively. The dark blue curve is the sensitivity curve for the LISA mission assuming a mission duration of $5$ years \citep{Robson_19}, while the brown, red, and green lines represent the $n=2$ harmonic of equal-mass, $10^{7}~{\rm M_{\astrosun}}$, $10^{6}~{\rm M_{\astrosun}}$, and $10^{5}~{\rm M_{\astrosun}}$ binaries, respectively. The dotted, dashed, and dash-dotted lines indicate systems initialized with various semi-major axes. We denote systems initialized with $e_0=0.45$ and $e_0=0.7$ with darker and lighter lines, respectively. The circles mark the point at which the binary reaches the eccentricity floor of $e=0.25$ determined in \autoref{ssec:cavity_req}. The triangles, squares, and stars, mark 1 month, 1 week, and 1 day before merger in the observed frame, respectively.
    }
    \label{fig:gw_strain}
\end{figure}

In the previous sections, we have outlined the necessary conditions for observing the effects of GR precession electromagnetically. While it may be challenging to identify the precession-induced modulations in the EM light-curve alone and to distinguish them from other sources of AGN variability, our system is also bright in gravitational waves. Thus, a GW detection followed up by observations in the X-ray and UV bands is likely to yield a much more robust determination of the mechanism driving the EM variability. The GR precession period is uniquely defined for $q$, $e$, $a$, and $M_{\rm bin}$, so if it is found imprinted on the EM light-curve, it can place excellent constraints on the binary parameters in combination with GW template searches. In \autoref{fig:gw_strain} we illustrate the observability of our systems in GWs by initializing binaries with $e$ and $a$ values that satisfy our EM observability constraints and then calculating their characteristic strain $h_c(f)$ and eccentricity $e$ as they shrink and circularize. 

Namely, we calculate the characteristic amplitude or ``strain" of the binary for the $n^{\rm{th}}$ harmonic
\begin{equation}
    \frac{1}{\pi D}\sqrt{2\frac{\left( G\mathcal{M} \right)^{5/3} \left( 2\pi  \right)^{2/3}  g_n(e)}{3F(e)\left(\left(1+z\right)f\right)^{1/3}c^3}}
\end{equation}
where

\begin{equation}
\begin{aligned}
    g_{n}(e) & = \frac{n^{4}}{32}  {} \left[ \lbrace J_{n-2}(ne) - 2eJ_{n-1}(ne) + \frac{2}{n}J_{n}(ne) \right. \\
    & + 2e J_{n+1}(ne) -J_{n+2}(ne)\rbrace ^{2} + \frac{4}{3n^2}J_{n}^{2}(ne)\\
    & + \left( 1-e^2 \right) \lbrace J_{n-2}(ne) \\
    & - 2J_n(ne)+ J_{n+2}(ne) \rbrace^{2} \left. \vphantom{\frac{1^{4}}{1}} \!\!\right],
\end{aligned}
\end{equation}

and
\begin{equation}
\begin{aligned}
    F(e) =  \sum_{n=1}^{\infty} g(n,e) = \frac{1+ \frac{73}{24}e^2 + \frac{37}{96}e^4 }{ \left( 1- e^2\right)^{7/2}},
\end{aligned}
\end{equation}
and $D$ is the luminosity distance to the source, 
\begin{equation}
\mathcal{M} \equiv M_{\rm{bin}}q^{3/5}\left(1+q\right)^{-6/5}
\end{equation}
is the chirp mass, and 
\begin{equation}
f \equiv n f_{\rm{bin}}\left(1+z\right)^{-1}
\end{equation}
is the observed harmonic frequency~\citep{peters_63,Barack_2004,ennoki_nagashima,huerta}.

In particular, \autoref{fig:gw_strain} displays various binary systems' strain and eccentricity evolution spanning mass, redshift, initial eccentricity, and initial semi-major axis. We only show the $n=2$ harmonics since these provide the highest strain in the parameter space displayed.

We find that at $z=1$ all our systems are loud enough to be detected by LISA and enter the LISA band just before they circularize to $e=0.25$, the eccentricity floor we determined in \autoref{ssec:cavity_req}. While more massive, initially less eccentric ($e_0$), and wider ($a_0$) binaries enter the LISA band just before reaching this eccentricity, the remainder of our systems at $z=1$ are well within the band before they reach $e=0.25$, i.e they are producing EM precession signals while being loud in LISA. A wide parameter space of $e_0$, $a_0$, and $M_{\rm{bin}}$ falls into this category at $z=1$, spanning nearly three orders of magnitude in $M_{\rm{bin}}$ and up to $1000~R_S$ in $a_0$. We see a similar situation at $z=3$, where most of our chosen systems produce an EM precession signal in the LISA band. However, due to the higher redshift, the strains are weaker, meaning that these systems produce the EM precession signal for less time in the LISA band.

While \autoref{fig:gw_strain} only depicts a number of selected systems we note that the trends it displays can be extrapolated. The average strain increases monotonically with $M_{\rm{bin}}$ and with decreasing $z$. We find that (i) systems with $M_{\rm{bin}}<10^{3}~M_{\astrosun}$ at $z=1$ are no longer detectable by LISA and (ii) the parameter space for systems producing a detectable EM precession signal while also detectable by LISA is greatly diminished at $z\geq 6$. Further, we note that an increase in $a_0$ shifts the strain curve towards lower frequencies, while a decrease in $e_0$ flattens the strain curve and shifts the eccentricity floor towards lower frequencies.

Though we only provide a sparse sampling of systems, the fact that a wide variety of binaries satisfy both EM and GW detection criteria through $z=3$ is encouraging for the prospect of identifying such sources in LISA and then following them up with EM observations to detect their EM precession signal.

\section{Summary and conclusions}\label{sec:conclusions}

In this paper, we presented results from two-dimensional hydrodynamic simulations of general relativistic precessing equal-mass binaries in circumbinary disks. Our results not only have implications for the broad hydrodynamics of binaries in circumbinary disks, but they detail a new periodic EM signal that could be useful for multi-messenger searches for MBHBs. We list our main results below.

\begin{enumerate}
    \item{We find that the apsidal precession of eccentric binaries ($e \gtrsim 0.25$) in CBDs creates a strong quasi-sinusoidal modulation in both the binary's net accretion rate and in the accretion rate ratio of the individual binary components. The period of this modulation is the time it takes for the binary to precess by $\pi$ ($\tau_{\rm{prec}}$).}
    
    \item{The X-ray and UV light-curves for eccentric, apsidally precessing binaries show corresponding quasi-sinusoidal modulations with a period of $\tau_{\rm{prec}}$. This behavior contributes a low-frequency peak in the system's periodogram that is second in strength only to the peak at the orbital frequency.}

    \item{Eccentric, apsidally precessing, massive ($10^5 - 10^7 ~{\rm M_{\astrosun}}$) binary black holes at $z \lesssim 3$ emit GWs whose harmonics are loud in LISA while simultaneously exhibiting this periodic EM signal.}
    
    \item{The X-ray and UV light-curves for eccentric binaries exhibit three flares per orbit with distinct hydrodynamic correlates. Apsidally precessing binaries uniquely display a periodic modulation in the amplitude and time separations of such flares.}

    \item{The optical light-curves for eccentric, apsidally precessing binaries do not share the accretion rate's quasi-sinusoidal behavior. However, they display a periodic modulation in the timing of the orbital-frequency flare.}

    \item{The cavities around eccentric, apsidally precessing binaries experience periodic modulations in their eccentricities on the timescale $\tau_{\rm{Prec}}$. The amplitude of the modulation is anti-correlated with the binary precession rate, suggesting that the mechanism for the change in cavity eccentricity is stream-cavity collisions. }

    \item{Near-circular binaries exhibit significant cavity precession and, as a result, their accretion rate and light-curve signatures exhibit strong modulations with a frequency matching the beat frequency between the precession frequencies of the binary and the cavity.}

    \item{The accretion behavior of a MBHB system can be well encapsulated by tracking the distances from the BH to the closest segment of the cavity ($r_{\rm{min}}$). $r_{\rm{min}_1}/r_{\rm{min}_2}$ tracks $\lambda \equiv \dot{M}_2/\dot{M}_1$, while the amplitude of $\dot{M}_1$ and $\dot{M}_2$ are dependent upon the values of $r_{\rm{min}_1}$ and $r_{\rm{min}_2}$.
    }
     
\end{enumerate}

In future work, we plan to expand our current analysis to a larger set of binary parameters, running simulations with a broader range of eccentricities, precession rates, and mass ratios to better understand the processes at play. We also plan to further study the shape and time-evolution of the cavity in this broad range of parameter space so that the EM signals from near-circular binaries can be generically determined and, thus, included as candidates. Furthermore, we plan to calculate the effect of gas torques (in isothermal and gamma-law equations of state) and understand how apsidal precession of the binary would affect disk-induced binary evolution, such as the equilibrium values of the eccentricity and semi-major axis \citep{zrake_eqbm_ecc,duffell_dorazio_2020,siwek_orbevol}.

Finally, we are interested in determining the light-curves of the system when viewed along different lines of sight since preferential accretion suggests one mini-disk is brighter than the other, which would enhance a Doppler signal when not viewed face-on.

We conclude by noting that an apsidal precession of the binary imprints an additional periodic modulation in the X-ray and UV light-curves on the precession timescale of $\tau_{\rm{Prec}}$. This EM signature of precession may be observable for lower-mass MBHBs ($10^{5} - 10^{7} ~M_{\astrosun}$) that have a significant eccentricity ($e \gtrsim 0.25$). With LISA triggers of sources up to $z\approx3$ we may be able to follow up on EM signals for a significant population of MBHBs---breaking degeneracies between the chirp mass, $e$, $a$, and other parameters. This novel multi-messenger signal has the potential to greatly refine our understanding of the eccentric binary black hole population, which hitherto has been poorly constrained by GWs alone.

\section*{Acknowledgments}

We thank Daniel D'Orazio for valuable discussions. SD acknowledges financial support from the Gates-Cambridge Trust. ZH acknowledges financial support from NASA grants 80NSSC22K0822 and 80NSSC24K0440 and NSF grant AST-2006176. JD is supported by a Joint Columbia/Flatiron Postdoctoral Fellowship. Research at the Flatiron Institute is supported by the Simons Foundation. JZ acknowledges financial support from NASA grant 80NSSC24K0440. AM acknowledges financial support from NASA ATP grant 80NSSC22K0822. We also acknowledge computing resources from Columbia University's Shared Research Computing Facility project, which is supported by NIH Research Facility Improvement Grant 1G20RR030893-01, and associated funds from the New York Empire State Development, Division of Science Technology and Innovation (NYSTAR) Contract C090171, both awarded April 15, 2010. This work made use of NASA-ADS. {\it Software}: {\tt Sailfish} \citep{ryan_sailfish}, {\tt python} \citep{travis2007,jarrod2011}, {\tt scipy} \citep{jones2001}, {\tt numpy} \citep{walt2011}, {\tt matplotlib} \citep{hunter2007}

\section*{Data availability}
The data underlying this article will be shared on reasonable request to the corresponding author.


\bibliographystyle{mnras}
\bibliography{main}

\end{document}